\documentclass[aps,twocolumn,amsmath,amssymb,preprintnumbers]{revtex4}
\usepackage{amsmath} \usepackage{amsfonts} \usepackage{amssymb}
\usepackage{bbm}
\usepackage{epsfig}
\usepackage{graphics}
\usepackage{graphicx}
\newcommand{\be}{\begin{equation}}
\newcommand{\ee}{\end{equation}}
\newcommand{\ba}{\begin{eqnarray}}
\newcommand{\ea}{\end{eqnarray}}
\newcommand{\nn}{\nonumber}
\newcommand{\kr}{\rangle}
\newcommand{\kl}{\langle}

\newcommand{\mn}{_{\mu\nu}}
\newcommand{\B}{_{B-L}}
\newcommand{\A}{\bar\alpha^2}
\newcommand{\Aa}{\tilde\alpha^2}

\begin{document}

\title[ ]{Variable gravity Universe}

\author{C. Wetterich}
\affiliation{Institut  f\"ur Theoretische Physik\\
Universit\"at Heidelberg\\
Philosophenweg 16, D-69120 Heidelberg}

\begin{abstract}
For variable gravity models the strength of gravity, as measured by Newton's ``constant'' or the Planck mass, depends on the value of a scalar field, the cosmon. 
We discuss two simple four-parameter models with a quadratic or constant cosmon potential. 
They are compatible with all presently available cosmological observations, including inflation. The inflaton and the scalar field of quintessence are the same cosmon field. Dark Energy constitutes a small, almost constant fraction of the energy density during the radiation and matter dominated epochs (Early Dark Energy). In the present epoch we witness a transition to a new Dark Energy dominated epoch. Our  models are free of a big bang singularity. The stability of solutions generates an arrow of time. Our picture of the Universe is unusual, with a shrinking or static scale factor, while the masses of particles increase and the size of atoms shrinks. The evolution of the universe can be very slow for all cosmological epochs including inflation, with typical time scale $10^{10}$ yr, and in sharp contrast to the usual big bang picture.  The map to the equivalent Einstein frame with constant particle masses and expanding scale factor  can be singular at the big bang.
\end{abstract}

\maketitle

\section{Introduction}
\label{Introduction}

Simple models of variable gravity give rise to realistic cosmology, describing inflation and the present transition to Dark Energy domination by the same scalar field. 
In these models gravity is modified, the Planck mass being proportional to the cosmon field $\chi$, at least asymptotically for large $\chi$. 
During the cosmological evolution the value of $\chi$ increases, thus reaching the asymptotic regime during or after the inflationary period. 

We present two simple models, one with a quadratic cosmon potential (A) and the other with a constant potential (B) or cosmological constant. 
The kinetic term of the cosmon makes a crossover from a positive constant for small $\chi$ to a negative constant close to the conformal value or stability bound for large $\chi$. Both models involve only three parameters in the scalar-graviton sector, plus an additional cosmologically relevant parameter which quantifies the growth of neutrino masses. This suffices for a realistic cosmology, compatible with all present observations. Models with such a simple structure are subject to interesting precision tests by forthcoming observations.

Besides the conceptual advantage of a unified description of all cosmological epochs we consider the simplicity of our models as an important novel feature of our approach. With two parameters fixed by the present fraction of Dark Energy and the amplitude of primordial density fluctuations it remains to be explored by future observations if the remaining two parameters are sufficient to account for all cosmological observations, involving the physics of primordial density fluctuations as well as present dynamical Dark Energy. Furthermore, variable gravity offers new intriguing pictures of the history of our universe, being in the past cold, evolving slowly and without a big bang singularity. This new view may influence research on the ``beginning'' of our universe. 

For a quadratic cosmon potential the Universe shrinks during radiation- and matter-domination, while for the constant potential the radiation epoch has a flat static Minkowski geometry. 
Both models lead to the same predictions of all observables for the radiation-, matter- and Dark Energy-dominated periods. 
All observable quantities resemble closely the ones in the standard $\Lambda$CDM-model. 
The main distinctive features are the presence of Early Dark Energy and the clumping of the cosmological neutrino background. 
The two models differ for the inflationary epoch, predicting for the density fluctuations a spectral index and tensor to scalar ratio $n=0.97,~r=0.13$ (quadratic potential) 
or a typical range $n=0.94-0.955$, $r=0.03-0.08$ (constant potential).

The asymptotic regime for $\chi \to \infty$ corresponds to the approach to a fixed point with the associated dilatation or scale symmetry. 
In this regime the electron and nucleon masses are proportional to the dynamical Planck mass $\chi$. 
These particle masses are therefore induced by the spontaneous breaking of scale symmetry for a nonzero $\chi$. 
Close to the fixed point two deviations from exact scale symmetry remain important. 
The first concerns the mass scales present in the cosmon potential. 
They induce a time varying mass for the cosmon, which otherwise would be an exact Goldstone boson. 
The second is a scaling of neutrino masses not proportional to $\chi$, as induced by scale violations in the evolution of a heavy singlet field which enters with inverse proportionality in the seesaw mechanism. These ingredients are sufficient to generate an overall cosmology compatible with observation for all times after the end of inflation. The inflationary epoch is followed by radiation- and matter-domination, while the 
present cosmological epoch is a transition to a Dark Energy dominated epoch.  As compared to the dynamical Planck mass $\chi$ all scale violating effects are tiny in the asymptotic regime, explaining the tiny value of the present Dark Energy density.

For the early epoch of inflation the violation of scale symmetry is no longer a small effect.  It determines the basic properties of inflation. In order to obtain a realistic setting with acceptable properties of the generated density fluctuations the scale violation in the scalar kinetic term plays an important role. The observed values of the spectral index and the amplitudes of scalar and tensor fluctuations impose constraints on the $\chi$-dependence of the kinetic term. These constraints are of qualitative nature, while the detailed form of the kinetic term does not matter. No fine tuning of parameters is necessary.

For our models the radiation- and matter-dominated epochs have an unusual geometry of spacetime. For model (A) the scale factor of the Universe is shrinking rather than expanding. The usual redshift due to the expansion is replaced by an increase of particle masses with increasing $\chi$. Light from distant galaxies has been emitted when the value of $\chi$ was smaller than its present value. Hence the electron and nucleon masses have been smaller, resulting in smaller frequencies. While traveling to us, the radiation has subsequently been blueshifted rather than redshifted. During the radiation dominated epoch the temperature was much smaller than the present $2.7^\circ$K and increasing. Particle masses $m_p$ increase even faster, however. The decrease of the ratio $T/m_p$ triggers cosmic events as nucleosynthesis or the emission of the cosmic microwave background (CMB) radiation. The picture of the past is a ``cold universe''. Furthermore, the characteristic time scale for the cosmic evolution is given for 
all cosmological epochs by the present inverse Hubble parameter of the order of $10^{10}$ yr. This includes inflation and the approach to the ``big bang'' for $t\to-\infty$. We may call this picture the ``slow cold universe''.

In model (B) the Universe expands in the matter-dominated epoch, although with a rate different from the usual one. For the radiation period the geometry is flat static Minkowski space and the temperature is constant.  The usual redshift is then entirely due to the increase of particle masses,  with a simple linear increase $\chi=c_\chi t$ in flat space. In this model the Dark Energy density is the same for all times, from inflation to now, $\rho_h=(2\cdot 10^{-3}$eV$)^4$.

Only dimensionless quantities are observable. Concerning the geometry of the Universe this involves the ratio of distances between galaxies divided by some fixed length, say the meter, which is in turn related to the size of atoms. Thus a shrinking atom size can be equivalent  to an expanding distance between galaxies. An increasing  mass of electrons $m_e\sim\chi$ is directly reflected  in a shrinking size of atoms $\sim\chi^{-1}$. Our picture of a shrinking Universe can be mapped to the usual picture of an expanding Universe by a field transformation of the metric.  

We present explicit solutions which remain regular for arbitrary values of time $t$. 
No big bang singularity is encountered for these solutions. (This holds even in the absence of an inflationary epoch. Realistic cosmology requires an inflationary epoch, however.) Despite the unusual geometry, the radiation and matter epochs lead to the same predictions for observables as in standard cosmology, except for small deviations due to the presence of a small almost constant fraction of Early Dark Energy. In particular, the results of nucleosynthesis are the standard ones. The clock provided by the Hubble expansion in the standard description is now replaced by a clock associated to the increasing value of $\chi$. The compatibility with standard cosmology is most easily seen by a Weyl scaling to the Einstein frame.

The idea that Newton's ``constant'' may be dynamical has a long history, going back to Dirac \cite{Dirac} and Jordan \cite{Jordan1,Jordan2}. In most early approaches the particle masses have been considered as constants \cite{Jo3,BD}. Then the variation of the ratio between Planck mass and nucleon mass - in our notation $\chi/m_n$ - is not compatible with observational bounds. For this reason models with otherwise interesting cosmological aspects \cite{OB,Fo} have not been considered as realistic. The situation changes radically \cite{CW1,CW3} if particle masses also scale proportional to $\chi$, i.e. $m_n(\chi)\sim\chi$, such that the ratio between Planck mass and nucleon mass, $\chi/m_n(\chi)$, remains constant. In this case all bounds on the time variation of fundamental constants and apparent violations of the equivalence principle are obeyed. 
The idea that the expansion of the Universe may be replaced by an increase of masses has been pursued earlier \cite{Na1,Na2,Na3}. These early models are, however,  not compatible with cosmological observations as nucleosynthesis or the cosmic microwave background. In contrast, the models proposed in this paper are consistent with all present observations.

If no parameter with dimension of a mass or length appears in the quantum effective action the theory is dilatation invariant or scale invariant \cite{Pen,Fu,BBPP}. The cosmology of such models has been investigated in refs. \cite{CW3,Fu}. After a Weyl scaling these models become in the Einstein frame standard gravity theories with a cosmological constant coupled to a massless dilaton without a potential \cite{CW3}. The dilaton is the Goldstone boson of spontaneously broken dilatation symmetry. With rather arbitrary initial conditions it settles to a fixed value in very early cosmology due to Hubble damping. For the observable cosmology of the radiation or matter epochs the dilaton plays no role - it is not generating a dynamical Dark Energy. 

Dynamical Dark Energy or quintessence  has been predicted \cite{CW3} from a setting where explicit scale symmetry breaking still plays a residual role. For example, a cosmological constant is associated to an explicit mass scale. This ``dilatation anomaly'' induces after Weyl scaling to the Einstein frame an exponentially decreasing potential for the scalar field. The scalar field becomes a pseudo-Goldstone boson, the cosmon, with a tiny mass that vanishes asymptotically as the role of the dilatation anomaly becomes less and less important. Our model (B) is similar in  spirit to the models with a cosmological constant in refs. \cite{CW1,CW3}. A cosmological constant may also arise as an integration constant in unimodular gravity without explicit scale symmetry breaking in the effective action. Concerning dynamical Dark Energy this yields the same cosmology as explicit symmetry breaking \cite{SZ1,SZ2,BS}.

Models with explicit dilatation symmetry breaking involve two types of mass scales: the intrinsic scales that we may denote here collectively by $m$, and the scale of spontaneous scale symmetry breaking $\chi$. For $\chi\gg m$ many details concerning the intrinsic scales will become irrelevant. One typical scenario is a ``runaway cosmology'', where $\chi$ continues to increase for increasing time, such that a fixed point is approached. At any fixed point dilatation symmetry is exact - the memory of intrinsic scales is ``forgotten'', up to their appearance in wave function renormalizations in case of non-vanishing anomalous dimensions. At the exact fixed point all particle masses must scale precisely proportional to $\chi$, while renormalized dimensionless couplings as gauge and Yukawa couplings take constant values. We will assume in this paper that masses and couplings of the standard model of particle physics are sufficiently close to the fixed point for large values of $\chi$ such that corrections from 
explicit scale symmetry breaking can be neglected in this sector. We will not make this assumption for non-renormalizable operators in the standard model of particle physics, as the terms generating a mass for the neutrinos. Here heavy singlet fields which enter the neutrino mass with inverse proportionality may not have reached a fixed point scaling $\sim\chi$. This will lead to models of ``growing neutrino quintessence'' \cite{ABW,CWNEU}. 

For the potential $V(\chi)$ of the cosmon a generalized ``fixed point'' is reached whenever $V(\chi)$ increases less fast than $\chi^4$ for $\chi\to\infty$. In this case, the dimensionless
 ratio $V(\chi)/\chi^4$ vanishes for $\chi\to\infty$, such that the cosmological constant in the Einstein frame vanishes asymptotically \cite{CW1,CW3,CW2}. Similarly, the squared cosmon mass in units of the Planck mass vanishes asymptotically and the cosmon becomes the massless dilaton. Obviously, in the other limit $\chi\lesssim m$ the intrinsic mass scales play a crucial role. This will be the case for the inflationary period in our models.

An important  new aspect of the present paper concerns the close link of the cosmon to inflation. 
During the inflationary epoch the energy density of the Universe is dominated by the potential energy of a scalar field, the inflaton, causing a very rapid expansion \cite{Inf1,Inf1A,Inf2,Inf3,Inf4,Inf5}. We link this inflationary phase to quintessence in late cosmology, where the scalar field again plays an important role \cite{CW3,RP,CW2,QU1,QU2,QU3,QU4,QU5}.
The cosmon can play the role of the inflaton, realizing a scenario of ``cosmon inflation'' \cite{CI}. (For other ideas relating Dark Energy and inflation see ref. \cite{PV,BM}.)
In order to obtain a realistic inflationary epoch it is sufficient that the coefficient of the kinetic term in the action, the ``kinetial'' $K(\chi)$, changes from a large positive value for small $\chi$ to a negative value for large $\chi$. Negative values remain compatible with stability due to the mixing between $\chi$ and the scalar component of the metric,  provided $K$ is larger than the conformal value $-6$. For our models the tight observational bounds on Early Dark Energy \cite{A2a,A2b,Re,A2c,A2d,PL} require that $K+6$ is small for large $\chi$. (A field dependence of the kinetial has been invoked previously in order to induce a crossover from matter domination to Dark Energy domination \cite{HebW,CWcoupl,CWcross}.)

The value of the cosmon field increases during inflation, typically from $\chi\ll m$ for the early stages of inflation to $\chi\gg m$ at the end of inflation. The intrinsic scale $m$ therefore plays an important role during the inflationary epoch. One may view the physics of inflation as the process of transition to the asymptotic regime during which the information about intrinsic mass scales is lost to a large extent. The role of the intrinsic scales $m$ for the properties of the density fluctuations created during inflation depends on the value of $x=\chi^2/m^2$ at the time when the associated length scales cross the horizon. 

In model (A) with a quadratic cosmon potential realistic inflation requires that $x$ is already much larger than one at horizon crossing. The intrinsic scale that appears in the slow roll parameters is related to the decrease of $K$ that is necessary to end inflation. We assume that $K$ decreases in the relevant region as $c/x$. Horizon crossing in the regime of large $x$ is realized for small $c$. For $c\to0$ one finds ``universal relations'' $\epsilon=\eta=1/2N$, with $N$ the number of $e$-foldings before the end of inflation at which the scale of the relevant fluctuations crosses the horizon. For $N=60$ this yields a spectral index $n=0.97$ and a tensor to scalar ratio $r=0.13$. The amplitude of the density fluctuations involves, in addition, the scale characterizing the size of the potential. Bounds on $r$ have been derived from Planck data \cite{PL} within the $\Lambda$CDM-model. The compatibility  of  model (A) with observations has to be checked by including in the analysis the predicted Early Dark 
Energy and the varying neutrino mass. 

Model (B) has a flat cosmon potential and an effective Planck mass $\sqrt{\chi^2+m^2}$ (in contrast to $\chi$ for model (A)). This form of the effective action is suggested by a recent functional renormalization analysis of a fixed point in dilaton quantum gravity \cite{HPRW}. For the decrease of $K$ we assume the same form as for model (A). In this model realistic density fluctuations are obtained in the regime of small $x$. 
Other mass scales besides the one governing the decrease of $K$ can play a role for the slow roll parameters. Typically, tensor amplitudes $r$ between $0.03$ and $0.08$ are found for a spectral index $n$ in the range between $0.94$ and $0.955$. 

The second theme of the present paper concerns an overall view of the different cosmological epochs in the ``Jordan frame'' where the Planck- and particle masses depend on $\chi$. In this scheme dilatation symmetry is realized by a multiplicative rescaling of the metric and scalar field. It is directly connected to the absence of parameters with dimension of mass or length. We will see that in the Jordan frame the cosmological solution has no big bang singularity. (In this paper we associate the ``Jordan frame'' with the choice of fields for which scale transformations are defined by a multiplicative rescaling of fields. In our approach both the Planck mass and all charged particle masses are proportional to a scalar field $\chi$ in the limit of large $\chi$. Our definition therefore differs from a definition of the Jordan frame as the one for which particle masses are constant.)

It is possible to use a  redefinition of the metric (conformal transformation or Weyl scaling \cite{Weyl}) in order to transform our model of variable gravity to the ``Einstein frame'' with a fixed Planck mass. The Einstein frame is advantageous for practical computations of observables since in this frame both the gravitational constant and the masses of nucleons or electrons have fixed values. For a given quantum effective action  the Jordan- and Einstein-frames are completely equivalent for all observables \cite{CW1,CW3}. This will allow us to perform the quantitative computations for the inflationary period in the familiar Einstein frame. Also the later radiation-, matter- and Dark Energy-dominated epochs are conveniently described in the Einstein frame where the field equations have the usual form.

Despite the practical advantages of the Einstein frame several important features of cosmology are understood easier in the Jordan frame. This concerns the issue of scale symmetry and its spontaneous breaking, together with the associated discussion of naturalness of cosmological equations in a quantum world \cite{CWEXP}. Furthermore, our models correspond to a very simple setting in the Jordan frame, while they would perhaps look contrived in a view from the Einstein frame. Perhaps most important in our context is the absence of a big bang singularity in the Jordan frame. For this reason we follow here the main cosmological epochs directly in the Jordan frame. In model (A) with a quadratic cosmon potential the evolution of the scale factor is given by a sequence of four de Sitter solutions for which $\chi$ increases exponentially. In the first inflationary epoch the Universe expands, while for radiation and matter domination it shrinks \cite{CWU}. Finally, the growth of the neutrino masses triggers the 
transition to a fourth Dark Energy -dominated epoch during which the Universe expands again.

In model (B) with a constant cosmon potential the radiation- and matter-dominated epochs show a linear increase of $\chi,~\chi\sim\sqrt{\bar\lambda_c} t$, with $\bar\lambda_c$ the cosmological constant. The early stages of inflation feature an almost exponential growth of $\chi$, while the later stages turn to a linear growth. Correspondingly the Hubble parameter is almost constant during the early stages of inflation, while it decreases $\sim t^{-1}$ towards the end of inflation. Most remarkably, the Hubble parameter vanishes for the radiation dominated epoch. For radiation domination the geometry is described by flat static Minkowski space! This turns over later to a scaling $H=1/(3t)$ for the matter dominated epoch. 

The map between the Jordan and the Einstein frame can be considered as a coordinate change in field space. (It is not a coordinate transformation of general relativity.) This explains  why models with a big bang singularity in the Einstein frame can be free of singularities in the Jordan frame. For our two models we may consider the Jordan frame as a choice of ``field coordinates'' that is free of singularities. The often discussed big bang singularity turns out to be a singularity of a coordinate transformation in field space, rather than having a physical meaning!

The present paper is organized as follows. We display general field equations for variable gravity cosmology in sect. \ref{Cosmon field equations}. Sects. \ref{Quadratic cosmon potential}-\ref{Cosmon inflation} deal with a quadratic cosmon potential, model (A). In sect. \ref{Quadratic cosmon potential} we specify the model, which uses for the $\chi$-dependent Planck mass the simple form $M_p(\chi)=\chi$, together with a quadratic cosmon potential. In sect. \ref{De Sitter solutions} we present the four de Sitter solutions which describe the four cosmological epochs in the limit of a constant kinetial $K$. In sect. \ref{Stability} we show that the de Sitter solutions with increasing $\chi$ are stable with respect to neighboring solutions as time increases, while they are unstable for decreasing time. This defines an arrow of time. A combination of the scalar field and the scale factor can be used as a cosmological clock. Sect. \ref{Numerical solutions} solves the cosmological equations numerically. We 
find that the solutions with constant $K$ are a very good approximation except for short transition periods. We also 
discuss bounce solutions where $\chi$ first increases and subsequently turns to the increase characteristic for the inflationary epoch.

In sect. \ref{Rescaled coordinates and fields} we bring our model to a picture closer to the standard description of cosmology. For temporal and spatial distances that are measured in units of the decreasing Planck length the Universe shows the usual expansion of the modified scale factor. A Weyl scaling to the Einstein frame results in an exponential cosmon potential. For this potential the inflationary epoch is still influenced by a field dependent kinetial. For the later radiation and matter dominated epochs the kinetial becomes almost constant and the exponential potential provides for a cosmic attractor (``tracker'') solution with Early Dark Energy. A growing neutrino mass finally stops the time evolution of the cosmon such that the potential acts as a cosmological constant. 

The field transformation to the Einstein frame becomes singular for $\chi\to 0$. This explains the appearance of a big bang singularity in the Einstein frame even though no such singularity is present in the field-coordinates of the Jordan frame. The Einstein frame is used in order to give a detailed discussion of the inflationary epoch in sect. \ref{Cosmon inflation}. 

Sects. \ref{Flat cosmon potential} and \ref{Cosmon inflation for flat cosmon potential} discuss our second model (B) with a constant potential. In sect. \ref{Flat cosmon potential} we discuss the various cosmological epochs in the Jordan frame. A more detailed discussion of the inflationary epoch in the Einstein frame is presented in sect. \ref{Cosmon inflation for flat cosmon potential}. We conclude in sect. \ref{Conclusions}.

\section{Cosmon field equations}
\label{Cosmon field equations}

The cosmological field equations can be derived by variation of the quantum effective action $\Gamma$. This includes already all effects of quantum fluctuations. (We do not attempt here an explicit computation of the quantum fluctuations, but rather assume that they result in a simple structure of $\Gamma$. No additional fluctuation effects should be added to $\Gamma$.) We consider a general form of the effective action for a scalar field $\chi$ - the cosmon - coupled to gravity
\ba\label{1}
\Gamma=\int d^4x\sqrt{g}\left\{-\frac12 F(\chi)R+\frac12K(\chi)\partial^\mu\chi\partial_\mu\chi+V(\chi)\right\}.
\ea
This is the most general form containing up to two derivatives which is allowed by diffeomorphism symmetry. (We assume that possible higher derivative  terms play a subleading role.) 

The effective action \eqref{1} remains rather generic and it is perhaps no surprise that realistic cosmologies can be obtained for a suitable choice of the three functions $F,K$ and $V$. Our aim will be the discussion of simple models. For this purpose we will later mainly concentrate on two models with a particularly simple action, namely
\be\label{A1}
(A):\quad F(\chi)=\chi^2~,~V(\chi)=\mu^2\chi^2,
\ee
and
\be\label{2A}
(B):\quad F(\chi)=\chi^2+m^2~,~V(\chi)=\bar\lambda_c.
\ee
The detailed dynamics of both models will be specified by the ``kinetial'' $K(\chi)$. We will follow a simple setting where $K$ interpolates between two different constant values for $\chi\to 0$ and $\chi\to\infty$. The present section discusses the general case with three functions $F,V$ and $K$.

An effective action of the type (1) has  been found in the context of ``higher dimensional inflation'' \cite{Inf5,HI1,HI2,HI3,HI4,HI5}, which describes inflation as the separation of the length scale of three space-like dimensions from the one of additional ``internal'' dimensions. By dimensional reduction such models are equivalent to four-dimensional inflation models, typically with constant $K,~F=\xi\chi^2$ and $V$ a polynomial of $\chi$. After a Weyl scaling this has introduced \cite{HI1} an exponential potential for the inflaton. Exponential potentials have later been widely studied in the context of power law inflation \cite{PLI1,PLI2,PLI3}.

Models of quintessence based on the action (1) are often called ``extended quintessence'' if $F$ deviates from a constant. They may be regarded as a simple form of modified gravity and have been studied widely \cite{CW1,CW3,EQ1,EQ2,EQ3,EQ4,EQ5}.
If particle masses do not depend on $\chi$ a Weyl scaling to the Einstein frame leads to an equivalent model  with standard gravity where the particle masses depend on the scalar field \cite{CW1,CW3}. These are models of ``coupled quintessence'' \cite{CW2,CQ1,CQ2,CQ2a,CQ3,CQ4,CQ5,CQ6,CQ7,CQ8}. For constant particle masses (in the Jordan frame), however, the coupling is universal and therefore  severely restricted by bounds on time varying couplings \cite{CW1,CWcoupl}. Since strong bounds can be found from nucleosynthesis \cite{NS1,NS2,NS3,NS4,NS5} such a coupling must be tiny since this time.  (A possible alternative for a large scalar  mass is the ``chameleon effect'',  whereby fundamental couplings depend on density and not on time \cite{DDC1,DDC2,DDC3,DDC4}.)
We note that $f(R)$ theories can also be cast into the form (1) \cite{FR1,FR2,FR3,FR4,FR5,FR6,FR7,FR8,FR9}, such that the same remarks apply. In the present paper we assume that the masses of charged particles are proportional to $\chi$, at least  for the cosmology  after inflation. In this case the cosmon coupling  to nucleons or electrons vanishes in the Einstein frame \cite{CW1,CW3}, and all observational constraints on time varying couplings or apparent violations of the equivalence principle are obeyed.

The particular emphasis of this paper concerns the simultaneous description of inflation and late quintessence by  a single scalar field. We will see that this can be realized if $K$ changes sign, with positive values for small $\chi$ and negative values for large $\chi$. This prevents us from setting $K=1$ by a simple field redefinition of the scalar field. We therefore consider the most general effective action (1) which contains at most two derivatives.

The scalar field equation derived from eq. \eqref{1} reads
\be\label{A2a}
-D_\mu(K\partial^\mu\chi) +\frac12\frac{\partial K}{\partial \chi}\partial^\mu\chi\partial_\mu\chi=-\frac{\partial V}{\partial\chi}+\frac12\frac{\partial F}{\partial\chi}R+q_{\chi},
\ee
and the gravitational field equation is given by 
\ba\label{A3}
&&F(R\mn-\frac12 Rg\mn)+D^2 Fg\mn-D_\mu D_\nu F\\
&&\quad +\frac12 K\partial^\rho\chi\partial_\rho\chi g\mn-K\partial_\mu\chi\partial_\nu\chi+V g\mn=T\mn,\nn
\ea
with $D_\mu$ the covariant derivative and $D^2=D_\mu D^\mu$. Here $T\mn$ is the energy momentum tensor of matter or radiation and $q_\chi$ differs from zero if particle masses depend on $\chi$. Insertion of the scalar field equation \eqref{A2a} into the Bianchi identity $D_\nu(R_\mu{^\nu}-\frac12 R\delta^\nu_\mu)=0$ relates $q_\chi$ and $T\mn$ \cite{CW1}. 

For a homogenous and isotropic Universe (and for vanishing spatial curvature) the field equations take the form \cite{CW1}
\ba\label{3}
K(\ddot{\chi}&+&3H\dot{\chi})+\frac12\frac{\partial K}{\partial \chi}\dot{\chi}^2= 
-\frac{\partial V}{\partial \chi}+\frac12 \frac{\partial F}{\partial\chi}R
+q_\chi,\\
FR&=&F(12H^2+6\dot{H})\label{6}\\
&=&4V-\left(K+6\frac{\partial F}{\partial\chi^2}\right)\dot{\chi}^2\nn\\
&&-6\frac{\partial F}{\partial\chi^2}(\ddot{\chi}+3H\dot{\chi})\chi-12\frac{\partial^2 F}{(\partial\chi^2)^2}
\chi^2\dot{\chi}^2-T^\mu_\mu,\nn\\
&&\hspace{-1.1cm} F(R_{00}-\frac12 R g_{00})=3FH^2\label{6A}\\
&&=V+\frac12K\dot{\chi}^2-6\frac{\partial F}{\partial\chi^2}H\chi\dot{\chi}+T_{00}.\nn
\ea
As usual, we denote the scale factor in the Robertson-Walker metric by $a(t)$ and $H=\partial_t\ln a$. 

The general consistency relation between $q_\chi, T_{00}=\rho$ and $T_{ij}=p\delta_{ij}$ reads
\be\label{8X}
\dot\rho+3H(\rho+p)+q_\chi\dot\chi=0.
\ee
(This can be verified easily by extracting $q_\chi$ from eq. \eqref{3}, $\rho$ from eq. \eqref{6A} and $p$ by combining eqs. \eqref{6} and \eqref{6A}.) 
For an ideal fluid of particles with a $\chi$-dependent mass $m_p(\chi)$ the explicit form of $q_\chi$ is given by
\be\label{8Y}
q_\chi=-\frac{\partial\ln m_p}{\partial\chi}(\rho-3p).
\ee
In particular, for $m_p(\chi)\sim\chi$ and $\rho-3p=m_p n_p$,
with $n_p$ the number density of particles, eq. \eqref{8Y} reads
\be\label{8Z}
q_\chi=-\frac{\rho-3p}{\chi}=-\frac{m_p}{\chi}n_p.
\ee
This will be the form assumed for charged particles, at least for large values of $\chi$. For neutrinos  we will assume a different form of $m_\nu(\chi)$ which multiplies effectively the r.h.s. of eq. \eqref{8Z} by a factor $(2\tilde\gamma+1)$, cf. sect. III.
For several species of particles with different masses $q_\chi$ is a linear superposition of the individual contributions.
Due to the variation of the mass the energy momentum tensor of the fluid is not separately conserved if $\dot\chi\not= 0$.

We may insert eq. \eqref{6} into eq. \eqref{3} and use the variable
\be\label{A5}
s=\ln \left(\frac{\chi}{m}\right),
\ee
with $m$ a fixed mass scale. This yields
\ba\label{A6}
&&\left[ K+6\frac{\chi^2}{F}\left(\frac{\partial F}{\partial\chi^2}\right)^2\right] 
(\ddot{s}+3H\dot{s})+
\left[K\left(1+\frac{\chi^2}{F}\frac{\partial F}{\partial\chi^2}\right)\right.\nn\\
&&\quad \left.+\frac{\chi}{2} \frac{\partial K}{\partial\chi}
+12\frac{\chi^2}{F}\left(\frac{\partial F}{\partial\chi^2}\right)^2+12
\frac{\chi^4}{F}\frac{\partial F}{\partial\chi^2}\frac{\partial^2 F}{(\partial\chi^2)^2}\right]\dot{s}^2\nn\\
&&\qquad =4\frac{\partial F}{\partial\chi^2}\frac VF-2\frac{\partial V}{\partial \chi^2}+\frac{q_\chi}{\chi}
-\frac{\partial F}{\partial\chi^2}\frac{T^\mu_\mu}{F},
\ea
where $\chi=m\exp(s)$ has to be inserted. For a determination of $H$ one uses eq. \eqref{6A}
\be\label{A7}
H^2=\frac{1}{3F}\left\{V+\frac12\chi^2 K\dot{s}^2-6\chi^2\frac{\partial F}{\partial\chi^2}H\dot{s}+T_{00}
\right\}.
\ee
This quadratic equation has typically two solutions which express $H$ as a function of $s,\dot{s}$ and $T_{00}=\rho$. Insertion into eq. \eqref{A6} yields a second order non-linear differential equation for $s$. In the presence of matter and radiation this has to be complemented by appropriate equations for $T\mn$ and $q_\chi$. 

\section{Quadratic cosmon potential}
\label{Quadratic cosmon potential}

Let us now concentrate on the simple class of models $(A),F=\chi^2,V=\mu^2\chi^2$. A given model in this class still needs the specification of the kinetial $K$. Realistic cosmologies can be obtained if $K$ is positive and sufficiently large for the values of $\chi$ relevant for the inflationary epoch, while it takes negative values close to the stability bound $(K=-6)$ for the large values of $\chi$ that are reached for the subsequent radiation, -matter- and Dark Energy-dominated epochs. The overall cosmological picture only depends on this qualitative behavior of the kinetial $K$. 

Quantitative issues require, however, a more detailed specification. We will mainly use in this work an interpolation between two constants, one for small $\chi$ and the other for large $\chi$. This involves three parameters, namely the limits $K_0=K(\chi\to 0),~K_\infty=K(\chi\to\infty)$ and the scale $m$ that characterizes the transition between these two limits. The details of the interpolation are less important, but have to be specified for numerical estimates. For model $(A)$ we take the form used in ref. \cite{CI,CWU}, 
\be\label{2}
K(\chi)=\frac{4}{\tilde \alpha^2}\frac{m^2}{m^2+\chi^2}+\frac{4}{\alpha^2}\frac{\chi^2}{m^2+\chi^2}-6.
\ee
The mixing of scalar degrees of freedom in $\chi$ and $g_{\mu\nu}$ leads to a stable theory since $K>-6$. (The special value $K=-6$ is the ``conformal point''. Note that for $F=\chi^2$ stability does not require $K>0$, but rather $K>-6$). The two limiting constants are $K+6=4/\tilde\alpha^2$ for $\chi^2\ll m^2$ and $K+6=4/\alpha^2$ for
$\chi^2\gg m^2$. Compatibility with observations in late cosmology (bounds an early dark energy) requires $\alpha\gtrsim 10$, while a realistic inflationary period in early cosmology can be realized for $\tilde \alpha\lesssim 0.02$. 

The present value of $\chi$ can be associated with the reduced Planck mass $M=2.44\cdot 10^{27}$eV, while the present value of $V=\mu^2\chi^2$ accounts for the dark energy density, such that 
\be\label{15AX}
\mu\approx 2\cdot 10^{-33}\text{eV}. 
\ee
Our model differs from a Brans-Dicke theory \cite{BD} by three important ingredients: the presence of a potential $V=\mu^2\chi^2$, the $\chi$-dependence of $K$ and, most important, the scaling of the masses of nucleons, charged leptons, gauge bosons and Higgs scalars proportional to $\chi$ \cite{CW1}.

For the choice \eqref{A1} the field equation \eqref{A6} simplifies considerably,
\be\label{A9}
(K+6)(\ddot{s}+3H\dot{s}+2\dot{s}^2)+\frac{\chi}{2}\frac{\partial K}{\partial\chi}\dot{s}^2=2\mu^2+g,
\ee
with 
\be\label{12AB}
\quad g=\frac{q_\chi}{\chi}-\frac{T^\mu_\mu}{\chi^2}.
\ee
The Hubble parameter is determined by eq. \eqref{A7}, 
\be\label{A10}
3(H+\dot{s})^2=\mu^2+\frac{K+6}{2}\dot{s}^2+\frac{\rho}{m^2}e^{-2s}.
\ee
The two solutions of eq. \eqref{A10},
\be\label{A11}
H+\dot{s}=\pm \sqrt{\frac{\mu^2}{3}+\frac{K+6}{6}\dot{s}^2+\frac{\rho}{3m^2}e^{-2s}}
\ee
are related to each other by time reversal. We observe that for $\rho\geq 0,K>-6$ the expression under the root is always positive. Unless these conditions are violated the sign of the root in eq. \eqref{A11} remains the same for all $t$. The combination $y=\ln a+s$ either monotonically increases or monotonically decreases. 

Since for a time reflection invariant model (as the present one) the two time directions are equivalent we can take the positive sign in eq. \eqref{A11} without loss of generality, resulting in the evolution equation 
\ba\label{A12}
&&(K+6)\ddot{s}=2\mu^2\nn\\
&&\qquad +(K+6)\left[\dot{s}^2-\sqrt{3}\dot{s}
\sqrt{\mu^2+\frac{K+6}{2}\dot{s}^2+\frac{\rho}{m^2}e^{-2s}}\right]\nn\\
&&\qquad -\frac{1}{2}\frac{\partial K}{\partial s}\dot{s}^2+g.
\ea
For radiation one has $T^\mu_\mu=q_\chi=0$, while for a fluid  consisting of particles with mass $\sim\chi$ one finds $T^\mu_\mu=\chi q_\chi=-(\rho-3p)$. For both cases one has $g=0$, such that effects of radiation and matter only enter via the term $\sim\rho$ in the square root. For constant $K$ this is also the only term which depends explicitly on $s$. Thus for $\rho=0$, $K=$const one has a shift symmetry $s\to s~+$ const. In this case eq. \eqref{A12} reduces to a first order differential equation for $\dot{s}$. 

We may also ask if the value $K=-6$ can ever be reached by a solution of eq. \eqref{A12} with $g=0$. This would require $(\partial K/\partial s)\dot{s}^2=4\mu^2$ and therefore $\partial K/\partial s>0$. On the other hand, if $K=-6$ is to be reached for an increasing $s$, starting from values $K>-6$, one would need $\partial K/\partial s<0$. One concludes that $K=-6$ cannot be reached in this case. This statement is independent of the precise choice of the kinetial $K(\chi)$ even if the function $K(\chi)$ crosses the value $-6$ for some $\chi_c$, $K(\chi_c)=-6$. It holds provided $(\partial K/\partial \chi)(\chi_c)<0$. Then the dynamics forbid that a solution $\chi(t)$ reaches the value $\chi_c$.

\section{De Sitter solutions}
\label{De Sitter solutions}

It is instructive to study simple explicit solutions which obtain in the limit of constant $K$. They are good approximations if $K$ varies sufficiently slowly such that 
\be\label{B1a}
\left|
\chi\frac{\partial \ln (K+6)}{\partial \chi}\right|=\left|
\frac{\partial\ln (K+6)}{\partial s}\right|\ll 1.
\ee
For the choice \eqref{2} this holds for both limits $\chi\to 0$ and $\chi\to\infty$. In sect. \ref{Numerical solutions} we will check the validity of this approximation. 

For constant $K$ the field equation \eqref{A12} has solutions where the geometry is for all times $t$ a de-Sitter space with constant $H$, while the effective Planck mass increases exponentially
\be\label{5}
H=b\mu~,~\chi=\chi_0\exp(c\mu t),
\ee
such that 
\be\label{B2a}
\dot{s}=c\mu~,~\ddot{s}=0.
\ee
This holds for pure scalar field cosmologies $(\rho=0)$ as well as in the presence of radiation or matter, provided that the energy density scales as
\be\label{B3a}
T_{00}=\rho=\bar\rho\mu^2\chi^2,
\ee
with constant $\bar \rho$. For $g=0$ eq. \eqref{A12} reduces then to an algebraic equation
\be\label{B4a}
2+(K+6)\left(c^2-c\sqrt{3+\frac{3(K+6)}{2}c^2+3\bar\rho}\right)=0,
\ee
implying 
\be\label{B5a}
c^2=\frac{3(K+6)\left(\sqrt{1+\frac{6K+28}{3K+18}\bar\rho+\bar\rho^2}-\bar\rho\right)-3K-14}
{(K+6)(3K+16)}
\ee
The Hubble parameter follows from eq. \eqref{A11} 
\be\label{22A}
b=-c+\sqrt{\frac13+\frac{K+6}{6}c^2+\frac{\bar\rho}{3}}.
\ee
A second solution obtains by time reversal, changing simultaneously the sign of $b$ and $c$.

The overall evolution of the Universe can be understood qualitatively as a sequence of de Sitter solutions. Different values of the proportionality constant $\bar\rho$ in eq. \eqref{B3a} for scalar-,  radiation- or matter-domination imply different values for $b$ and $c$. Also realistic Dark Energy can be described as a de Sitter solution. Varying particle masses result in this case in $g\not= 0$, modifying the algebraic equation for $b$ and $c$. Furthermore, realistic models require different constants $K$ for very early and for late cosmology. It is striking that $\bar\rho,b,c,g,K$ are all constants of order one. There is therefore only a unique scale $\mu\approx 2\cdot  10^{-33}$eV that governs the  time evolution for all epochs in the history of the Universe.  In our picture the size of the Hubble parameter is always given by the scale $\mu$. 

\medskip\noindent
{\bf {\em 1. Scalar domination.}}

Consider first solutions in the absence of radiation or matter, $\bar \rho=0$. Provided $(K+6)(3K+16)>0$ one has the solutions
\be\label{B6a}
c=\pm\frac{2}{\sqrt{(K+6)(3K+16)}}.
\ee
The two signs are related by time reflection and we may concentrate on the positive sign with increasing $\chi$ and $s$. For $K>-6$ this solution exists for a range of $K$ obeying
\be\label{B7a}
K>-\frac{16}{3}.
\ee
The values of the Hubble parameter corresponding to the solutions \eqref{B6a} follow from eq. \eqref{A11}
\ba\label{B8a}
b&=&\pm\sqrt{\frac13+\frac{K+6}{6}c^2}-c\nn\\
&=&\pm \frac{K+4}{\sqrt{(K+6)(3K+16)}}=\frac{K+4}{2}c.
\ea
The Universe is expanding $(b>0)$ for increasing $\chi(c>0)$ if $K\geq -4$, otherwise it is shrinking. The sign of the product 
\be\label{B9a}
bc=\frac{2(K+4)}{(K+6)(3K+16)}
\ee
is independent of the ``direction of time'', i.e. independent of the sign of the roots in eqs. \eqref{B6a} and \eqref{B8a}. (The signs in eqs. \eqref{B6a} and \eqref{B8a} are not independent, cf. eq. \eqref{A9}.)

\medskip\noindent
{\bf {\em 2. Asymptotic early cosmology.}}

We begin with scalar field dominated cosmology and assume $\tilde \alpha^2<2$ such that for $\chi\to 0$ the condition $K>-4$ is obeyed. Then scalar field dominated cosmology describes an exponentially expanding Universe with exponentially increasing effective Planck mass $\chi$. As long as constant $K$ remains a good approximation the solution \eqref{5}, \eqref{B6a}, \eqref{B8a} can perfectly describe the evolution of the Universe for all times, including $t\to-\infty$. This solution is completely regular, no singularity is encountered. Indeed, we can take for $t\to-\infty,\chi\to0$ the geometry of a de Sitter space for which the curvature tensor
\be\label{10A}
R_{\mu\nu\rho\sigma}=b^2\mu^2(g_{\mu\rho}g_{\nu\sigma}-g_{\mu\sigma}g_{\nu\rho})
\ee
and all its covariant derivatives are regular.

The ``big bang'' is now free of any singularities! The central ingredient why the usual singularity is avoided arises from the behavior of the effective Planck mass $\chi$: it approaches zero as $t\to-\infty$. From the point of view of the field equations \eqref{3} this is in no way problematic, even though the effective strength of gravity, characterized by the effective Newton-constant $G(\chi)=1/(8\pi\chi^2)$, diverges for $t\to-\infty$. 

As $\chi$ grows with increasing time the approximation of constant $K$ will no longer remain valid. In the region of $\chi$ around $m$ we will have to investigate solutions where the $\chi$-dependence of $K(\chi)$ is taken into account. For very large $\chi^2\gg m^2$ we may again use a constant $K$. For $2<\alpha^2<6$ we have again the solution \eqref{5}, \eqref{B6a}, \eqref{B8a}, but now with negative $b$ and therefore negative $H$. In this region the scale factor $a(t)$ decreases. For $\alpha^2>4$ we have further the solution \eqref{5}, \eqref{B11} with negative $H$. For $\tilde \alpha^2 <2$ and $\alpha^2>2$ a pure scalar field cosmology therefore describes a Universe where the scale factor $a(t)$ first increases exponentially, and subsequently decreases exponentially. For all times the effective Planck mass $\chi$ grows exponentially. 

\medskip\noindent
{\bf {\em 3. Inflation.}}

We will next show that the first stage of the evolution describes an inflationary Universe. After the end of inflation, entropy will be produced and the Universe is heated. For this evolution after the end of inflation we therefore need to add radiation and matter, such that a pure scalar cosmology is no longer valid. Nevertheless, the main picture of cosmology is a sequence of de Sitter spaces, the first with an expanding Universe (increasing $a(t)$) and subsequently with a shrinking Universe (decreasing $a(t)$).
The end of inflation will be triggered by an increase of the kinetial $K$. Even though de Sitter solutions remain a good approximation we will include the slow variation of $K(\chi)$ according to eq. \eqref{2}.

Let us take $\tilde \alpha \ll 1$. For the very early epoch with $\chi\ll m$ one has $K+4=4/\tilde\alpha^2-2\gg 1$, such that $b\gg c$. In this case we can neglect $\ddot{\chi}$ as compared to $3H\dot{\chi}$ in eq. \eqref{3}. This property is called the ``slow roll approximation'' for inflation. We may continue the slow roll approximation to  larger values of $\chi$. As long as $\chi^2/m^2\ll \alpha^2/\tilde \alpha^2$ we can neglect in eq. \eqref{2} the term $\sim \alpha^{-2}$ such that the evolution equations read in the slow roll approximation
\ba\label{10}
H^2&=&\frac{\mu^2}{3},\nn\\
\dot{\chi}&=&\frac{\tilde \alpha^2\mu\chi(m^2+\chi^2)}{\sqrt{3}(m^2-3\tilde\alpha^2\chi^2)}.
\ea
The slow roll approximation breaks down once $\dot{\chi}/\chi$ is roughly of the same order as $H$. We may define
\be\label{11}
\tilde\epsilon=\left(\frac{\dot{\chi}}{H\chi}\right)^2=
\left(\frac{\tilde \alpha^2(m^2+\chi^2)}{m^2-3\tilde\alpha^2\chi^2}\right)^2,
\ee
such that the slow roll period ends once $\tilde \epsilon$ is of the order one. (For large $K$ the criterion is rather $\tilde\epsilon\lesssim K$.) For $\chi^2/m^2=1/(4\tilde\alpha^2)$ one reaches $\tilde\epsilon=1$ and we conclude that the inflationary slow roll phase ends once $\chi$ reaches a value of this order of magnitude. The amplitude of density fluctuations is governed by the ratio of the potential over the fourth power of the effective Planck mass, $\mu^2/\chi^2$. For large values of $m^2/(\tilde \alpha^2\mu^2)$ the density fluctuations can be very small, as required for a realistic cosmology. We will present a more detailed quantitative description of the inflationary epoch in sect. \ref{Cosmon inflation}.

\medskip\noindent
{\bf {\em 4. Radiation domination.}}

For radiation the energy density scales $\rho_r\sim a^{-4}$. 
With this scaling  and for constant $K$ we find again  a de Sitter solution \eqref{5}, \eqref{B2a},  \eqref{B3a},  \eqref{B5a}, \eqref{22A}.
In order to realize eq. \eqref{B3a} with constant $\bar \rho_r$ the scalar field has to evolve as $\chi\sim a^{-2}$. This requires
\be\label{B10}
b=-\frac c2.
\ee
The scaling value $\bar\rho$ has to be adapted such that eq. \eqref{B5a} has a solution obeying the condition \eqref{B10}. One finds
\be\label{B11}
c=\frac{2}{\sqrt{K+6}}~,~b=-\frac{1}{\sqrt{K+6}}
\ee
with 
\be\label{B3}
\bar\rho_r=-3\frac{K+5}{K+6}.
\ee
Again, there exists also the time reversed solution with opposite signs of $b$ and $c$. 
For constant $K$ the exact de Sitter solution \eqref{B11},  \eqref{B3} remains regular for all finite $t$. For $t\to -\infty$ one has $\chi \to 0,\ a\to \infty$. We note that no big bang singularity occurs even within a setting equivalent to the Friedman Universe, i.e. without invoking an inflationary period.

A positive radiation density $\rho_r$ requires $K<-5$. We therefore consider eq. \eqref{2} with large $\chi^2/m^2$, e.g. $K+6=4/ \alpha^2$, such that a radiation dominated Universe requires
 $\alpha^2>4$. (For $K=-5$ the solution matches smoothly with scalar field dominated cosmology according to the solution \eqref{B6a}, \eqref{B8a}.) We can compute the fraction $\Omega_h$ of homogenous Dark Energy in the total energy density $\rho_c=\rho_r+\rho_h$, namely 
\ba\label{B4}
\rho_h&=&\mu^2\chi^2+\frac12 (K+6)\dot{\chi}^2,\nn\\
\Omega_h&=&\frac{\rho_h}{\rho_r+\rho_h}=\frac{1}{K+6}=\frac{4}{\alpha^2}.
\ea
Positive $\rho_r$ and $\rho_h$ requires $0\leq \Omega_h\leq 1$ which is obeyed for all $\alpha^2\geq 4$.
The fraction of Dark Energy $\Omega_h$ during nucleosynthesis cannot be too large without affecting the successful computation of element abundancies. The strong  bound $\alpha > 10$ from the CMB-limit for Early Dark Energy suppresses $\Omega_h$ sufficiently such that the role of Dark Energy during nucleosynthesis is minor.

The scenario of a shrinking radiation dominated Universe with increasing effective Planck mass looks rather unfamiliar and intriguing. The temperature scales $T\sim (\rho_r)^{1/4}\sim \chi^{1/2}$. Since $\chi$ is monotonically increasing the temperature during the radiation dominated epoch was much smaller than today's temperature of the CMB, and it was increasing. This contrasts with the standard big bang picture. However, the particle masses as the electron or nucleon masses were increasing $\sim \chi$, even faster than the temperature. As a result, the ratio $m_e/T$ was decreasing. Cosmic events as nucleosynthesis or the combination of protons and electrons to hydrogen are triggered when the ratio $T/m_e$ falls below a certain value.

We will see in sect. \ref{Rescaled coordinates and fields} that this scenario predicts actually the same observations as the standard picture of radiation dominated Universe with expanding scale factor and constant Planck mass. In particular, one recovers the same element abundancies from nucleosynthesis, up to small modifications from $\Omega_h$.

\medskip\noindent
{\bf {\em 5. Matter domination.}}

A realistic setting requires that the mass of the nucleon $m_n$ or the electron $m_e$ scale proportional to the growing Planck mass $\chi$. Otherwise the ratio $m_n/\chi$ would depend on time, violating the strict observational bounds. (Small deviations from this proportionality are allowed and could result in an observable time variation of fundamental constants and apparent violation of the equivalence principle \cite{CW1,Da1,DZ,TC,CWcoupl,UZ,Da2,Da3}.)  For the electron and the other charged leptons and quarks this can be achieved by an effective potential for the Higgs doublet $\tilde h$ for which the expectation value $\kl \tilde h\kr$ is proportional to $\chi$. An  example is
\be\label{C1a}
\tilde V_h=\frac12\lambda_h(\tilde h^\dagger\tilde h-\epsilon_h\chi^2)^2,
\ee
with constant $\epsilon_h,\lambda_h$.
Similarly, a scaling of the hadron masses proportional to $\chi$ is realized if the gauge couplings take fixed values at some scale $M_g$ (for example the unification scale of a grand unified theory), and if $M_g$ scales $\sim\chi$, similar to $\kl \tilde h\kr$. Then $\Lambda_{QCD}$ and therefore $m_n$ is proportional to $\chi$. In such a setting dilatation symmetry is only broken by the scale $\mu$ in the cosmon potential and by a non-trivial $\chi$-dependence of the dimensionless kinetial $K$. The latter is negligible for $\chi^2\gg m^2$. 

In this paper we will assume that dark matter consists of particles whose mass also scales $\sim\chi$. We will call the nucleons, electrons and dark matter particles collectively ``charged particles'', in distinction to the neutrinos.
As a consequence of the proportionality of particle masses to $\chi$ one finds for the charged particles an additional ``force" in eq. \eqref{3}, adding a term \cite{CW1}
\be\label{C1b}
q_\chi=-(\rho_m-3p_m)/\chi
\ee
on the right hand side. Also on the r.h.s of eqs. \eqref{6}, \eqref{6A} one has now to add terms $-T^\mu_\mu/\chi^2=(\rho_m-3p_m)/\chi^2$ and $T_{00}/\chi^2=\rho_m/\chi^2$, respectively. (Generalizations to a different form of $q_\chi$ for dark matter are  possible.)

For a conserved particle number the density $n$ is diluted as $n\sim a^{-3}$. Thus the energy density of a pressureless gas will scale $\sim\chi a^{-3}$. In the radiation dominated period we have found $\chi\sim a^{-2}$ such that $\rho_m\sim a^{-5}$. Since the Universe is shrinking during radiation domination the ratio $\rho_m/\rho_r\sim a^{-1}$ increases until $\rho_m$ becomes comparable to $\rho_r$ at some time. After this matter-radiation equality we can essentially neglect radiation and follow the evolution in a matter dominated Universe. 

For $\rho_m\sim\chi^2$ (and neglecting $p_m$) the additional terms on the r.h.s. of the field equations are constant (after dividing eq. \eqref{3} by $\chi$). Solutions of the type \eqref{5} are again possible, now with 
\ba\label{B5}
\rho_m&=&\bar \rho_m\mu^4\exp \big\{(-3b+c)\mu t\big\},\nn\\
\frac{\rho_m}{\chi^2}&=&\bar\rho_m\mu^2~,~3b+c=0.
\ea
With the addition of the corresponding terms on the r.h.s the field eqs. \eqref{3}, \eqref{6}, \eqref{6A} or eqs. \eqref{B5a}, \eqref{22A} are all obeyed for constant $K$ and
\ba\label{B6}
c=\sqrt{\frac{2}{K+6}},~\qquad~b=-\frac{1}{3}\sqrt{\frac{2}{K+6}}=-\frac13c,
\ea
with 
\be\label{B7}
\bar\rho_m=-\frac{2(3K+14)}{3(K+6)}.
\ee
This solution exists for $K<-14/3$ or $\alpha^2>3$. For this solution the Dark Energy fraction is given by
\be\label{38A}
\Omega_h=\frac{\rho_h}{\rho_m+\rho_h}=\frac{3(K+6)}{4}=\frac{3}{\alpha^2}.
\ee
For both the radiation and matter dominated epochs a constant fraction of the energy density corresponds to homogenous Dark Energy, similar to the ``tracking solution'' in ref. \cite{CW3,CW2}.

The matter and radiation dominated epochs can both be described with the same constant value for $K$, provided $K<-5$. The transition occurs by a change in the dominant component of $\rho$. The scalar dominated de Sitter solution \eqref{B6a}, \eqref{B8a} exists for $K>-16/3$. For the range $-16/3<K<-5$ all three solutions \eqref{B6a}, \eqref{B8a} or \eqref{B11}, \eqref{B3} or \eqref{B6}, \eqref{B7} exist simultaneously. One may start with scalar field domination and small $\rho_r,\rho_m$. In order to judge the fate of this solution we have to compare the term $\rho/\chi^2$ with $\mu^2$ in eq. \eqref{A12}. For $\rho\sim a^{-4}$ and $\chi\sim a^{(c/b)}$ one has 
\be\label{C2a}
\frac{\rho}{\chi^2}\sim a^\kappa~,~\kappa=-\left(\frac{2c}{b}+4\right). 
\ee
Inserting eq. \eqref{B8a} yields
\be\label{C3a}
\kappa=-\frac{4(K+5)}{K+4},
\ee
which is negative in this range. Since $a$ is decreasing for $K<-4~(b<0)$ the ratio $\rho/\chi^2$ increases until it becomes comparable to $\mu^2$. The scalar field dominated cosmology makes then a transition to radiation domination.

We will be interested in models where $K$ decreases for increasing $\chi$. Starting from positive values and decreasing to a value smaller than $-5$ various qualitative changes occur. In the early epoch the Universe expands for $K>-4$. Once $K$ crosses the value $-4$ the Universe starts shrinking. A second threshold is at $K=-5$ where $\kappa$ turns from a positive to a negative value. Now the radiation energy density $\rho_r$ can become important and induce a radiation dominated epoch. The matter energy density starts growing as compared to radiation as soon as particles become non-relativistic. This triggers the transition to the matter dominated epoch.

At this point cosmology is described by a sequence of three (approximate) de Sitter geometries, all with exponentially increasing $\chi$. For the first scalar dominated inflationary period the Universe is expanding, while it shrinks for the subsequent radiation and matter dominated epochs. The Hubble parameter $H=b\mu$ remains always of the same order of magnitude, changing sign, however, after the end of inflation. 

\medskip\noindent
{\bf {\em 6. Dark energy domination.}}

We may now discuss the present transition to a fourth dark energy dominated epoch. Let us suppose that the $\chi$-dependence of the masses of neutrinos differs from the one of the charged leptons or nucleons. Within the standard model of particle physics Majorana masses for the light left-handed neutrinos are generated by non-renormalizable interactions which violate lepton number. The relevant terms are quadratic in the Higgs doublet and involve the inverse of a heavy mass scale $M\B$. Typically, $M\B$ corresponds to the mass of singlet ``right handed'' neutrinos for the seesaw mechanism \cite{ss1,ss2,ss3}, or is a combination the mass of a heavy triplet with a cubic coupling for the cascade (or seesaw II) mechanism \cite{MCW1,MCW2,MCW3, MCW4,MCW5}. This is an important difference to the masses of charged particles which are proportional to the Higgs doublet and arise from renormalizable interactions. 

The proportionality of the electron mass to the variable Planck mass $\chi$ (or $F^{1/2}(\chi))$ is typically realized by the attraction of the ratio of doublet expectation value over the Planck mass to a fixed point as $\chi$ grows large \cite{CWcross,CWvaryingcouplings}. The same holds for dimensionless Yukawa couplings and gauge couplings, ensuring that the QCD scale $\Lambda_{QCD}$ scales also proportional to $\chi$. A fixed point entails dilatation symmetry for the quantum effective action. As a result, no mass scale appears in the effective potential \eqref{C1a} for the Higgs doublet. 

The singlet sector responsible for lepton number violation may not have settled at a fixed point, but rather slowly drift with $\chi$, e.g. 
\be\label{N1}
\frac{M_{B-L}(\chi)}{\chi}=F_{B-L}-G\B\ln\left(\frac{\chi^2}{\mu^2}\right).
\ee
For increasing $\chi$ the ratio $M\B(\chi)/\chi$ decreases. (This could explain why the scale $M\B$ is typically a few orders of magnitude below the Planck mass or some grand unified scale even though $F\B$ could be of the order one.) The masses of the light neutrinos 
\be\label{40A}
m_\nu\sim\frac{\tilde h^2}{M\B}\sim \frac{\epsilon_h\chi^2}{M\B(\chi)}
\ee
show an additional increase from this effect. 

We arrive then at a simple picture: for early cosmology the ratio of neutrino mass to the Planck mass has been much smaller than at present, due to the additional suppression factor $M_{B-L}(\chi_{\rm today})/(\chi_{\rm today}F_{B-L})$. Only once $\chi$ approaches the value where $M\B(\chi)/\chi$ is substantially smaller than $F\B$ the ratio of neutrino mass to the Planck mass reaches the present size. This has happened around the present cosmological epoch, and we can use this for determining a relation between $F\B$ and $G\B$. Denoting by $\chi_0$ the value of $\chi$ where $M\B(\chi_0)=0$ we have
\be\label{N2}
\frac{M\B(\chi)}{\chi}=G_{B-L}\ln\left(\frac{\chi^2_0}{\chi^2}\right).
\ee
The masses of the light neutrinos therefore show a scaling violation 
\be\label{N3}
m_\nu(\chi)=\frac{c_\nu\chi}{\ln\left(\frac{\chi^2_0}{\chi^2}\right)}.
\ee

In the approximation \eqref{N1} the neutrino mass diverges for $\chi\to\chi_0$, but this value is actually never reached. What is characteristic, however, is the strong dependence of $m_\nu(\chi)$ on $\chi$ as $\chi$ comes close to $\chi_0$. As a typical measure for this dependence we introduce
\be\label{N4}
\tilde \gamma(\chi)=\frac12\chi\frac{\partial}{\partial\chi}\ln\left(\frac{m_\nu(\chi)}{\chi}\right).
\ee
If neutrino masses would scale $\sim\chi$, similar to the electron mass, one would have $\tilde\gamma=0$, whereas for the behavior \eqref{N3} one has
\be\label{N5}
\tilde\gamma=\frac{1}{\ln\left(\frac{\chi^2_0}{\chi^2}\right)}.
\ee
In the relevant range of $\chi$ only somewhat smaller than $\chi_0$ one expects large values of $\tilde \gamma$. 

In order to gain some qualitative understanding we may first consider some constant value of $\tilde\gamma$. For constant $\tilde\gamma$ the neutrino mass scales as
\be\label{N6}
m_\nu=\bar c_\nu\chi^{2\tilde\gamma+1}.
\ee
As a consequence, the ratio of neutrino energy density $\rho_\nu$ as compared to the matter density $\rho_m$ will increase with time. The Universe will make a transition to a new scaling solution. We will see that this new epoch is dominated by dark energy associated to the cosmon potential. 

For $\chi$-dependent neutrino masses according to eq. \eqref{N4} or  \eqref{N6} the source on the r.h.s of the scalar field equation  \eqref{A2a} obeys 
\be\label{N7}
\chi q_\chi=-(2\tilde\gamma+1)(\rho_\nu-3 p_\nu). 
\ee
This term becomes important once neutrinos get non-relativistic, such that $p_\nu$ can be neglected. The new scaling solution obeys, with constant $\bar\rho_\nu$
\be\label{N8}
\frac{\rho_\nu}{\chi^2}=\bar\rho_\nu\mu^2.
\ee
Since $\rho_\nu$ scales $\sim\chi^{2\tilde\gamma+1}/a^3$ this requires 
\be\label{N9}
b=\frac13(2\tilde\gamma-1)c.
\ee
For $\tilde\gamma>1/2$ the scale factor is now again expanding. 

The algebraic equation from eq. \eqref{A10} remains unmodified
 \be\label{N10}
b+c=\sqrt{\frac13+\frac{K+6}{6}c^2+\frac{\bar\rho}{3}},
\ee
while the l.h.s. of eq. \eqref{B4a} receives an additional term from $g$ in eq. \eqref{A12}
\be\label{N11}
\frac{g}{\mu^2}=-\frac{2\tilde\gamma\rho_\nu}{\mu^2\chi^2}=-2\tilde\gamma\bar\rho_\nu.
\ee
This replaces in eq. \eqref{B4a} the first term $2$ by $2-2\tilde\gamma\bar\rho_\nu$. In eq. \eqref{A9} the ``force'' $g$ due to the increasing neutrino mass, $g=-2\tilde\gamma\bar\rho_\nu\mu^2$, counteracts the force $2\mu^2$ from the gradient of the potential. Eq. \eqref{A9} implies
\be\label{52A}
b=\frac{2(1-\tilde\gamma\bar\rho_\nu)}{3(K+6)c}-\frac{2c}{3},
\ee
and comparison with eq. \eqref{N9} yields
\be\label{52B}
\tilde\gamma\bar\rho_\nu=1-(K+6)c^2\left(\tilde\gamma+\frac12\right).
\ee
From eqs. \eqref{N9}, \eqref{N10} we finally obtain 
\be\label{52C}
c^2=\left(\frac{K+6}{2}+\frac{4\tilde\gamma(1+\tilde\gamma)}{3}\right)^{-1}
\ee
and therefore
\be\label{52D}
\bar\rho_\nu=\frac{8(1+\tilde\gamma)-6(K+6)}{8\tilde\gamma(1+\tilde\gamma)+3(K+6)}.
\ee

We may compute the ratio between Dark Energy and the neutrino energy density
\ba\label{52E}
\frac{\rho_h}{\rho_\nu}=\frac{\Omega_h}{\Omega_\nu}=
\frac{8\tilde\gamma(1+\tilde\gamma)+6(K+6)}{8(1+\tilde\gamma)-6(K+6)}=
\frac{\tilde\gamma(1+\tilde\gamma)+\frac{3}{\alpha^2}}{1+\tilde\gamma-\frac{3}{\alpha^2}},
\ea
or
\be\label{52EA}
\Omega_h=1-\frac{1}{1+\tilde\gamma}+\frac{3}{\alpha^2(1+\tilde\gamma)^2}.
\ee
For large $\alpha$ this yields the approximate ratio 
\be\label{52F}
\frac{\Omega_h}{\Omega_\nu}\approx \tilde\gamma ~,~\Omega_h\approx 1-\frac{1}{\tilde\gamma+1}.
\ee
We conclude that Dark Energy dominates for large $\tilde \gamma$. With $m_\nu$ the average neutrino mass ($m_\nu=\sum m_{\nu i}/3)$ and 
\be\label{52G}
\Omega_\nu=\frac{m_\nu}{16eV}
\ee
this yields for the Dark Energy density the well known \cite{ABW} formula 
\be\label{52H}
\rho^{1/4}_h=1.27\left(\frac{\tilde\gamma m_\nu}{eV}\right)^{1/4} 10^{-3}eV,
\ee
which relates Dark Energy quantitatively to the neutrino mass. 

In our context $\tilde\gamma$ is only an approximation. Taking the field-dependence of $\tilde\gamma$ into account we recover the growing quintessence scenario of ref. \cite{CWNEU}. We observe in eq. \eqref{52C} the relation 
\be\label{52I}
\lim_{\tilde\gamma\to\infty}c^2=\frac{\sqrt{3}}{2\tilde\gamma},
\ee
such that $\dot{\chi}/\chi\to 0$ for $\tilde\gamma\to\infty$. For a continuously increasing $\tilde\gamma$ the time evolution of the cosmon effectively stops before $\chi$ reaches the value $\chi_0$. 

\medskip\noindent
{\bf {\em 7. Intrinsic mass scales}}

The simplicity of the history of the Universe is striking for our model A with a quadratic cosmon potential. It is a sequence of four (approximate) de Sitter solutions. The Universe expands during the early and late scalar field dominated epochs, while it shrinks during radiation and matter domination. The overall size of the Hubble parameter and the curvature scalar is set for all four periods by the same intrinsic scale $\mu$. 

For a fixed dimensionless ratio $m/\mu$ the parameter $\mu$ is the only mass scale in our model. Its value is arbitrary. In particular, it is not related to the masses of nucleons or electrons that are induced by spontaneous dilatation symmetry breaking $\sim\chi$. (The ratio between the dynamical Fermi scale $\tilde h$ - given by the minimum of $\tilde V$ in eq. \eqref{C1a} - and the dynamical Planck mass $\chi$ involves a very small, so far unexplained, dimensionless coupling $\sqrt{\epsilon_h}$.) Nevertheless, for practical purposes it is useful to use units where today’s value of the Planck mass takes the standard value,
\be\label{52J}
\chi_{\rm today}=M=2.44\cdot 10^{27}{\rm eV}.
\ee
In these units one has
\be\label{52K}
\mu=2\cdot 10^{-33}{\rm eV}.
\ee
This follows by identifying the present value of the cosmon potential, $\mu^2\chi^2_{\rm today}$, with the present Dark energy density. 

The large ratio $M/\mu\approx 10^{60}$ is not a parameter of our model. It is a ``historical number'', simply reflecting how much time has elapsed since the moment when $\chi$ was equal to $\mu$. (This is similar to the value $\rho_m/\chi^4_{\rm today}\approx 10^{-120}$ which is also a result of the time evolution of the matter energy density $\rho_m$ and of $\chi$.) The evolution of $\chi$ is exponential,
\ba\label{52L}
\chi(t)\approx \mu \exp (\bar c \mu t),
\ea
where we fix $t=0$ by the condition $\chi(0)=\mu$ and $\bar c$ is some appropriate average of $c$. Most of the increase of $\chi$ has happened during the radiation and matter dominated epochs where $\bar c\approx \alpha$. (For radiation domination one has $c=\alpha$, for matter domination $c=\alpha/\sqrt{2}$.) In units of $\mu^{-1}$ the present value of $t$ is not very large
\be\label{52M}
t_{\rm today}\approx \frac{60\ln 10}{\bar c}\mu^{-1}.
\ee
(For an order of magnitude estimate one may take $\mu t_{\rm today}\approx 100/\alpha\leq 10.)$ For this model the Universe is not extremely old in its natural time units. Only  the dynamical Planck mass has increased by a large exponential factor.

The ratio $m/\mu$ appearing in the kinetial \eqref{2} will be fixed by the amplitude of the density fluctuations generated during inflation, cf. sect. \ref{Cosmon inflation}. It depends on $\tilde \alpha$. For $\tilde\alpha\approx 10^{-3}$ one needs $m/\mu\approx 100$. We conclude that $\mu$ is indeed the only relevant mass scale. For all cosmological epochs the characteristic time scale of the evolution is given by $\mu^{-1}$, which is of the order $10^{10}$ yr. Our picture describes a ``slow universe'' for the whole evolution since infinite negative time.

\section{Stability and arrow of time}
\label{Stability}

In this section we investigate the stability of the various de Sitter solutions in the approximation of constant $K$. Our starting point is eq. \eqref{A12} with $g=0$, supplemented by a suitable equation for $\rho$. 

\medskip\noindent
{\bf {\em 1. Stability for scalar dominated epoch.}}

We begin with the scalar field dominated cosmology $(\rho=g=\partial K/\partial s=0$). 
In this case eq. \eqref{A12} involves $\ddot{s}$ and $\dot s$, while
 $s$ does not appear explicitly. Therefore,    eq. \eqref{A12} becomes a first order differential equation for $c(t)=\dot{s}(t)/\mu$,
\be\label{S1a}
\frac{\dot{c}}{\mu}=\frac{2}{K+6}+c^2-c
\sqrt{\frac32\big(2+(K+6)c^2}\big).
\ee
(In contrast to the exact de-Sitter solutions $c(t)$ depends now on cosmic time.) 

For small deviations of $c(t)$ from the constant $c$ as given by eq. \eqref{B6a},
\be\label{S1}
c(t)=\frac{2}{\sqrt{(K+6)(3K+16)}}+\delta(t),
\ee
one obtains the linearized equation
\ba\label{S2}
\dot{\delta}=-A_c\mu\delta~,~A_c=\sqrt{\frac{3K+16}{K+6}}.
\ea
The solution 
\be\label{S3}
\delta=\delta_0\exp (-A_c\mu t)
\ee
approaches the scaling solution as $t$ increases if $A_c>0$, i.e. $\delta(t\to\infty)\to 0$. This is indeed the case for the range $(K+6)(3K+16)>0$ for which the scaling solution exists. We conclude that the scaling solution with increasing $\chi$ is stable. It will be approached asymptotically by all solutions in its vicinity.

In contrast, the solution is unstable in the direction of negative time where $\chi$ decreases. For our time reversal invariant model the sign of time is a pure convention. We could also investigate the time reflected setting with negative constant $c$, i.e. $c(t)=-2\big((K+6)(3K+16)\big)^{-1/2}$, now with a positive sign of the root in eq. \eqref{S1a}. This will result in a change of sign for $A_c$ in eq. \eqref{S2}. Now the solution with increasing time and decreasing $\chi$ is unstable, while the direction of decreasing time and increasing $\chi$ is stable. We arrive at the important conclusion that the stability of the scaling solution singles out the direction of increasing $\chi$. 

Any particular cosmological solution exists for both directions of time, at least for a certain time interval. The properties of the solutions with positive root in eq. \eqref{A11} and the time reflected ones with negative root are, however, rather different if we try to extend them for $t\to\infty$. For the positive root the scaling solution is approached as $t\to\infty$. For given initial conditions at $t_{in}$ this solution exists for arbitrary $t\geq t_{in}$. In contrast, for the negative root one observes a divergence of $\dot{s}$ and $H$ at some critical time $t_c>t_{in}$. This behavior singles out an ``arrow of time''. We may associate positive time to the direction into which solutions can be extended to $t\to\infty$. 
For generic solutions this requirement
 fixes the sign of the root in eq. \eqref{A11} to be positive. The divergence of a generic solution occurs then in the negative time direction for some $t_c<t_{in}$. 

There is one exception to the generic behavior of solutions with negative root in eq. \eqref{A11}, namely the exact scaling solution \eqref{B6a}, \eqref{B8a} with negative $c$. If the initial condition at $t_{in}$ is given exactly by $\dot{s}(t_{in})=-2\mu\big((K+6)(3K+16)\big)^{-1/2}$ one finds $\chi$ decreasing to zero for $t\to\infty$. However, any difference $\epsilon$ in the initial condition
\be\label{S4}
\dot{s}(t_{in})=-\frac{2\mu}{\sqrt{(K+6)(3K+16)}}+\epsilon,
\ee
leads to a divergence at finite $t_c$. The smaller $\epsilon$, the smaller is the value of $\chi_c=\chi(t_c)$. We see that the characteristics of the solution depends now very sensitively on the precise initial condition, in sharp contrast to the vicinity of the scaling solution with increasing $\chi$. 

Let us now fix the choice of the time coordinate such that the sign of the root in eq. \eqref{A11} is positive. It is instructive to compare solutions with positive or negative $\dot{\chi}(t_{in})$. The solutions with increasing $\chi$ soon reach the close vicinity of the scaling solution \eqref{B6a}, \eqref{B8a} after a certain time. The memory of the precise initial conditions is rapidly lost. In contrast, a solution with decreasing $\chi,\dot{\chi}(t_{in})<0$, will generically reach a turning point where $\chi$ stops decreasing and increases subsequently. (We discuss these ``bounce solutions'' in more detail in the next section.) As time goes on, also these solutions reach the vicinity of the scaling solution with increasing $\chi$. 

\medskip\noindent
{\bf {\em 2. Cosmic clock and arrow of time.}}

We have already discussed in the previous section that the combination
\be\label{S5}
y=s+\ln a
\ee
either increases or decreases monotonically as long as $K>-6$ and $\rho\geq 0$. This quantity can be used for a definition of a physical time variable which is independent of the coordinate choice for $t$. The arrow of time can now be defined by the direction in which $y$ increases. In this direction the solutions approach asymptotically the scaling solution and can be continued to infinite time. For the scalar field dominated cosmology, as well as for radiation or matter domination, the asymptotic increase of $\chi$ and the increase of $y$ are correlated. Still, $y$ remains also increasing for all parts of a given solution, even for the epoch before the bounce where $\chi$ is decreasing. It is this monotonic behavior that singles out $y$ for a useful ``cosmological clock''. 

It is instructive to discuss our findings from the perspective of time reversal symmetry. Our model is time reflection invariant, but any given non-static solution is not. Thus time reversal symmetry is spontaneously broken for any given cosmological solution. This explains why the evolution equation \eqref{S2} for small deviations $\delta$ from the scaling solution is no longer time reversal invariant. It is this feature that permits to define an arrow of time by the direction in which the scaling solution is stable. The assignment of a positive or negative coordinate time $t$ remains arbitrary and a matter of convenience, as expected for spontaneous symmetry breaking of time reversal invariance. However, the direction of the ``physical time'' $y$ acquires as objective measurable meaning. The positive direction is the direction towards a stable scaling solution, while the negative direction makes the scaling solution unstable. The two directions have therefore different properties which can be measured. We 
will from now on fix the coordinate time $t$ such that it increases with increasing $y$. This fixes the positive sign of the square root in eq. \eqref{A11}.

\medskip\noindent
{\bf {\em 3. Predictivity and singularities.}}

Local solutions of eq. \eqref{S1a}  are characterized by one ``integration constant'' or ``initial condition''  $\dot s(t_{in})$. For generic initial conditions the solution becomes singular for some $t_c < t_{in}$, with the exception of the regular exact de Sitter solution for
\be\label{75A}
\dot s(t_{in})=\frac{2\mu}{\sqrt{(K+6)(3K+16)}}.
\ee
In view of our statement about the absence of a big bang singularity one may wonder if the presence  of nearby singular solutions  does not invalidate our argument.  We will show that the presence  of neighboring local diverging solutions is  actually a characteristic feature of regular solutions which define an arrow of time by their stability. 

Universally stable  solutions of the type \eqref{S3} permit a high degree of predictivity. Once a solution is in the vicinity of the stable solution at some time $t_1$, with small $\delta_1=\delta(t_1)$, eq. \eqref{S3} implies that for a later time $t_2$, with sufficiently  large $t_2-t_1\gg (A_c\mu)^{-1}$, the solution is exponentially  close to the scaling solution, i.e. $\delta (t_2)=\delta (t_1)\exp(-A_c\mu(t_2-t_1))$. Up to tiny corrections the solution is {\it predicted} to be the stable scaling solution for times around $t_2$. This prediction becomes exact for infinite $t_2-t_1$, e.g. for $t_1\to -\infty$. In other words, at $t_2$ the solution  has (almost) lost memory of possible initial conditions in the past.

If we associate  $t_{in}$ with $t_2$, and we assume that the equations for the scalar field dominated cosmology have been valid already for a large time interval  $t_2-t_1$ before $t_{in}$, the ``initial condition'' for  $\dot s(t_{in})$ is very severely restricted by the predictivity of the setting. If one violates this ``prediction'' and chooses at $t_{in}$ initial conditions not compatible with the small predicted value of $\delta_2$, one will typically be reminded of the inconsistency of this approach by a solution diverging in the time interval between $t_1$ and $t_2$. This argument extends qualitatively (with possible exceptions) to rather arbitrary  values  of $\delta(t_1)$ (not necessarily small). We conclude that a regular big bang cosmology simply predicts the solution to be the exact de Sitter solution. If other physics plays a role and the scalar dominated cosmology becomes insufficient  before a finite $t_1$, one still remains with  very tight restrictions for $\delta (t_2)$. This predictivity  
is directly  related to the loss of memory of the detailed physics at $t_1$.

\newpage

\noindent
{\bf {\em 4. Stability and arrow of time for radiation 

~domination.}}

The stability of the scaling solutions for increasing $y$ also holds for the radiation and matter dominated epochs.  In the presence of radiation or matter eq. \eqref{A12} becomes a second order differential equation. Its solution requires two initial conditions, $s(t_{in})$ and $\dot{s}(t_{in})$. Furthermore, we have the conservation equation for $\rho$ which requires to fix a third initial condition, namely $\rho(t_{in})$. Altogether, the generic solutions will therefore depend on three initial conditions, instead of only one for scalar field dominated cosmologies.

For the radiation dominated epoch the energy momentum tensor of radiation is conserved, resulting in the usual evolution for $\rho$ 
\be\label{S6}
\dot{\rho}=-4H\rho.
\ee
For eq. \eqref{A12} we only need the quantity $\rho/\chi^2=\rho e^{-2s}/m^2$. The evolution equation for this ratio reads 
\be\label{S7}
\partial_t\left(\frac{\rho}{\chi^2}\right)=-(4H+2\dot{s})\frac{\rho}{\chi^2}.
\ee
We conclude that the general solution of eq. \eqref{A12} actually requires only two initial conditions, $\dot{s}(t_{in})$ and $(\rho/\chi^2)(t_{in})$. The third initial condition, i.e. $\rho(t_{in})$, will not affect the stability analysis. In addition to $\delta$, as defined by eq. \eqref{S1}, we therefore introduce a second parameter $\gamma$ for deviations from the scaling solution, namely
\be\label{S8}
\frac{\rho}{\chi^2}=\mu^2(\bar\rho+\gamma),
\ee
with $\bar\rho$ given by the scaling solution \eqref{B3}. 

The linearized evolution equations for $\delta$ and $\gamma$ read 
\ba\label{S9}
\partial_t\gamma&=&-\mu(A_{\gamma\gamma}\gamma+A_{\gamma\delta}\delta),\nn\\
\partial_t\delta&=&-\mu(A_{\delta\gamma}\gamma+A_{\delta\delta}\delta),
\ea
with 
\ba\label{S10}
\begin{array}{lll}
A_{\gamma\gamma}=-\frac{2(K+5)}{\sqrt{K+6}}&,&A_{\gamma\delta}=-\frac{2(K+5)(2K+9)}{K+6},\\
A_{\delta\gamma}=1&,&A_{\delta\delta}=\frac{2K+11}{\sqrt{K+6}}.
\end{array}
\ea
Positive real parts of the eigenvalues of the $2\times 2$-matrix A correspond to stability. We find that both eigenvalues are real and positive for $-(81/16)\leq K\leq 5$, 
\be\label{S11}
\lambda_{1,2}=\frac{1}{2\sqrt{K+6}}(1\pm \sqrt{16 K+81}),
\ee
while they get complex for $K<-(81/16)$,
\be\label{S12}
\lambda_{1,2}=\frac{1}{2\sqrt{K+6}}(1\pm i\sqrt{-16K-81}).
\ee
For all $K<-5$ the scaling solution is approached for $t\to\infty$ and therefore stable for increasing time. The approach to the scaling solution is oscillatory for $K<-81/16$. For the negative time direction the scaling solution is unstable. One generically finds again a singularity at $t_c<t_{in}$, except for initial conditions corresponding to the exact scaling solution. 

It is instructive to compare the radiation dominated epoch for our model with constant $-6<K<-5$ with standard cosmology in the presence of radiation. For standard cosmology eq. \eqref{A11} is replaced by 
\be\label{S13}
H=\pm \sqrt{\frac{\rho}{3M^2}},
\ee
with $M$ the (fixed) reduced Planck mass. We may again fix positive time by the positive root in eq. \eqref{S13}. The general solution of the systems of equations \eqref{S13} and \eqref{S6} reads
\be\label{S14}
\rho=\frac{3M^4}{(2Mt+\epsilon_\rho)^2}.
\ee
For large $t\gg|\epsilon_\rho|/M$ it approaches the scaling solution $\rho=3M^2/(4t^2)$. This solution is stable towards positive $t$. 
For arbitrary $\epsilon_\rho$ it has a singularity at $t=-\epsilon_\rho/(2M)$.
The branch of the solution with increasing $\rho$ for $t<-\epsilon_\rho / (2M)$ corresponds to the time reversed solution.
One may define the positive time direction such that the Universe is expanding. 
(In a more general setting with spatial curvature this holds for an energy density smaller than the critical density, $\Omega<1$, such that $\ln a$ increases monotonically.) 
We observe that the physical time observable $y$ in our model corresponds to $\ln a$ in standard cosmology. 

We should recall that the stability analysis of this section assumes a spatially flat geometry. The instability of the radiation dominated Universe in case of non-zero spatial curvature (instability of $\Omega=1)$ is the same as usual. Realistic inflationary scenarios guarantee $\Omega$ extremely close to one at the beginning of radiation domination, such that this instability plays no role.

\section{Numerical solutions}
\label{Numerical solutions}

So far we have concentrated  on an analytic discussion of the properties of solutions with constant kinetial $K$. They reflect all  qualitative features of our model. It is, of course, also possible to solve numerically eq. \eqref{A12}, combined with a suitable equation for $\rho$. This can be done for an arbitrary kinetial $K(\chi)$ or $K(s)$. In this section we briefly  display solutions for the kinetial  \eqref{2} and $g=0,\rho=0$. Extensions to the later periods with $\rho>0$ are straightforward. However, except  possibly  for the very early stages of radiation domination, the kinetial $K$ is so close to the constant $K=4/\alpha^2-6$ that effects from the $\chi$-dependence of $K$ are  uninterestingly small.

\medskip\noindent
{\bf{\em 1. Inflationary solutions.}}

In  Fig.  1 we show the dependence of $s$ on cosmic time $t$ for three different initial conditions. 
(Parameters for the figures are $\mu=1,\ m=5,\ \bar\lambda_c=0,\ \alpha=10,\ \tilde\alpha=0.3$.)
After a short initial  period all curves rapidly approach  the asymptotic solution with a linear increase of $s(t)$. We have verified that this holds for arbitrary initial conditions as long as $s$ remains in the range where $K$ is sufficiently positive. This behavior illustrates the stability properties discussed in the preceding section.

In Fig. 2 we have extended the scaling solution to larger values of $t$. 
We show the Hubble parameter $H(t)/\mu$ and the derivative of $s,\ w(t)=\dot s(t)/\mu$. 
Both remain almost constant until the time when the kinetial is close to zero. 
Around $t=60$ the evolution of the scalar field  accelerates and the inflationary period ends when $\dot s$ grows large. 
The slow increase of $\dot s$ before the end of inflation reflects the change in the kinetial $K$. 
Despite this time-variation of $K$ the approximate  de Sitter solutions for constant $K$ remain valid with high accuracy. 
We also plot in Fig. 2 the values of $b(t)$ and $c(t)$ which are defined by eqs. \eqref{B6a} and \eqref{B8a}, inserting for $K$ the value of $K(t)=K(\chi(t))$. 
There is no optical distinction between $H(t)/\mu$ and $b(t)$ or $w(t)$ and $c(t)$, except for the very end  at $t\approx 60$ where differences around 10\% occur.

\begin{figure}[h!]
\includegraphics[width=\linewidth]{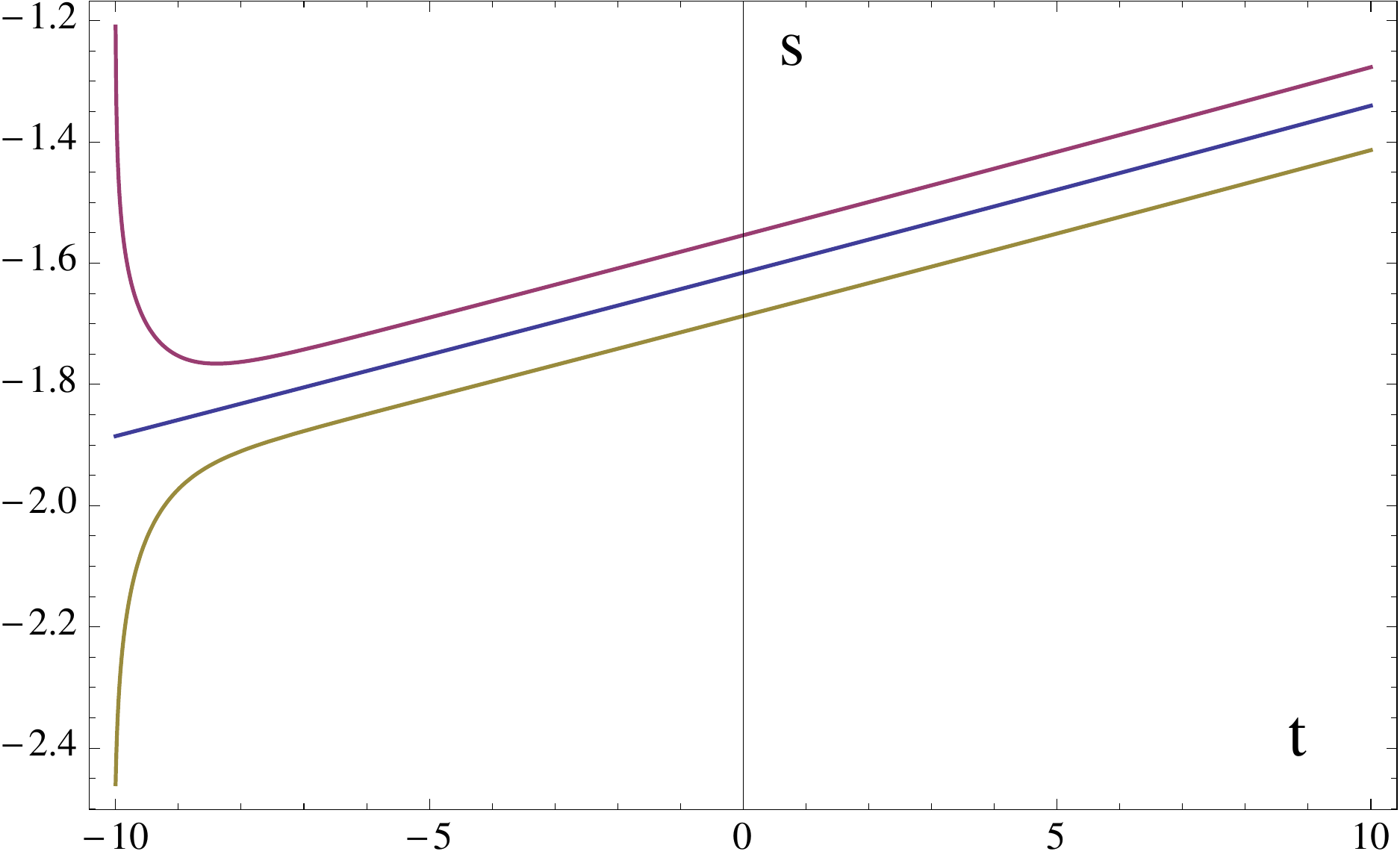}
\caption{Time dependence of scalar field $s(t)$ in early cosmology.}
\end{figure}

\begin{figure}[h!]
\includegraphics[width=\linewidth]{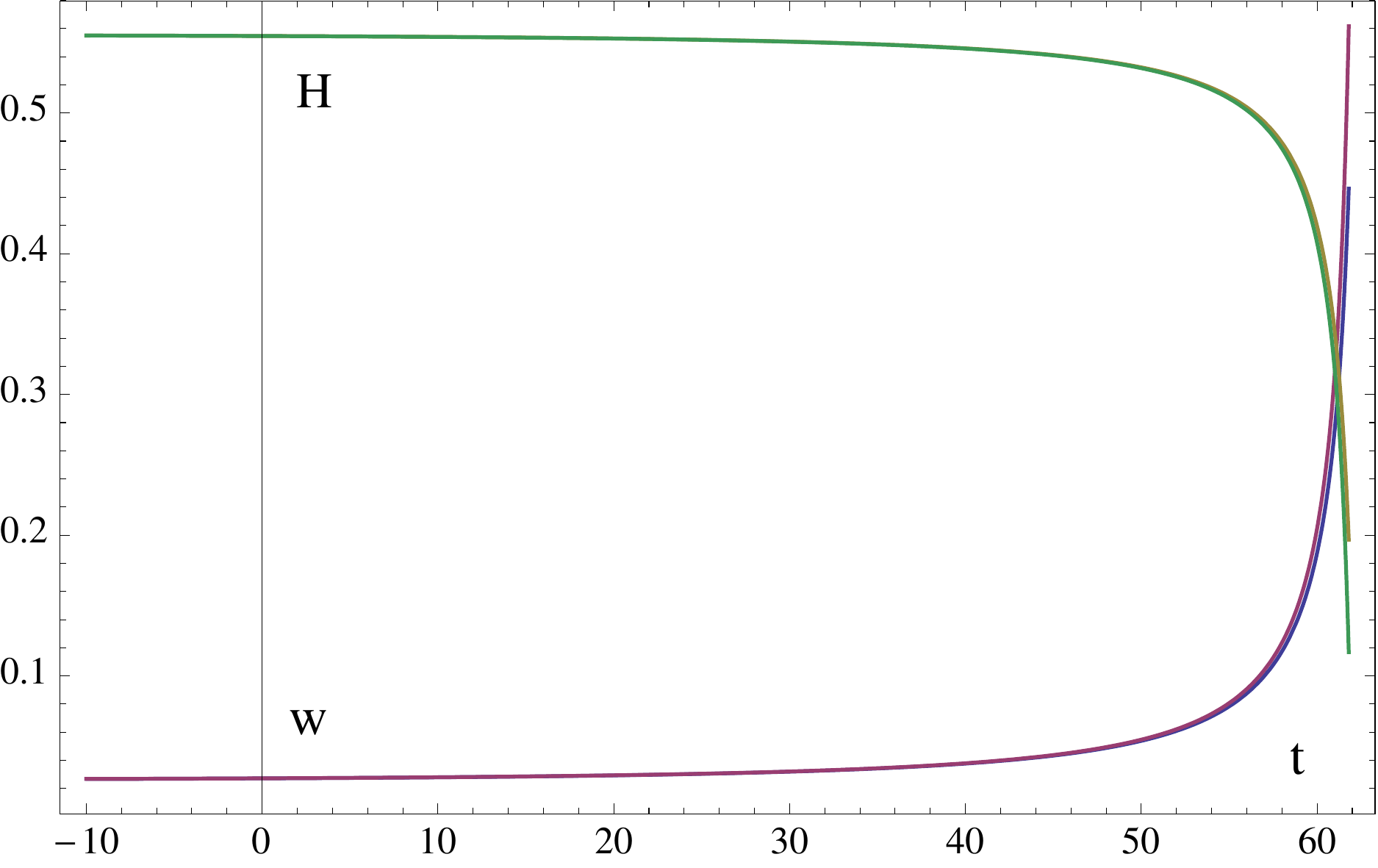}
\caption{Hubble parameter $H(t)/\mu$ and time derivative of scalar $w(t)=\dot s(t)/\mu$ for the inflationary period. We also display $b(t)$ and $c(t)$ which cannot be distinguished optically from $H(t)/\mu$ and $w(t)$.}
\end{figure}

\medskip\noindent
{\bf{\em 2. Bounce solutions.}}

The system of field equations \eqref{3}, \eqref{6}, \eqref{6A} admits regular bouncing solutions, whereby $\chi$ first decreases, stops, and increases subsequently. After the bounce it approaches the de Sitter solution \eqref{B6a}, \eqref{B8a}. 
This behavior is demonstrated by the upper curve in Fig. 1.
For an analytic discussion  we take again constant $K$.  Eq. \eqref{A9}  yields
\be\label{T1}
\ddot{s}+3H\dot{s}+2\dot{s}^2=\frac{2\mu^2}{K+6}=f,
\ee
while the Hubble parameter is extracted from eq. \eqref{A11}
\be\label{T2}
H=\pm\frac{\mu}{\sqrt{3}}\sqrt{1+\frac{\dot{s}^2}{f}}-\dot{s}.
\ee
(Measuring time in units of $\mu^{-1}$ we can put $\mu=1$.)

The constant positive force $f$ pushes $s$ towards large positive values. It is reduced by a type of friction force $-2\dot{s}^2$ which is the same for positive and negative $\dot{s}$. For positive $H$ and positive $\dot{s}$ the term $-3H\dot{s}$ acts as an additional damping force. For negative $H\dot{s}$, however, this term even enhances the constant force $f$. If we start a solution with some initial negative $\dot{s}$ and positive $H$ the combined action of $f-3H\dot{s}$ will bring $s$ to a stop, and $\dot{s}$ will become positive subsequently. (Near the turning point one has $\ddot{s}\approx f,H=\mu/\sqrt{3}$.) For positive $\dot{s}$ the damping will result in a saturation of $\dot{s}$ at a constant value $c$ given by eq. \eqref{B6a}. Starting initially with an expanding Universe and decreasing effective Planck mass $\chi$, the point $\chi=0$ will never be reached. Instead, there will be a turnaround to an increasing $\chi$.

\section{Rescaled coordinates and fields}
\label{Rescaled coordinates and fields}

In this section we map our model and its cosmological solutions to more familiar formulations of gravity coupled to scalars. We first discuss a rescaling of coordinates. It maps the solutions with a shrinking Universe to an expanding Universe. Second, we perform the field transformation from the Jordan frame to the Einstein frame. In the Einstein frame we will find quintessence scenarios with an exponential  potential for late time. Physical observables do not depend  on the frame.  We can therefore be sure that  our model reproduces the predictions of standard cosmology  for the radiation  and  matter dominated  phases, except  for small modifications due to the presence of Early Dark Energy.

\medskip\noindent
{\bf {\em 1. Rescaled coordinates.}}

The interpretation of cosmologies with a variable effective Planck mass becomes more familiar if we choose a different system for the coordinates. So far we have fixed time intervals $dt$ and comoving space intervals $dx^k$ in a Robertson-Walker metric. Instead, we may want to measure time in units of the inverse effective Planck mass $\chi^{-1}$, introducing the time intervals
\be\label{E1}
dt'=\frac{\chi(t)}{M}dt~,~t'(t)=\frac1M\int\limits^t_{t_0}\chi (\hat t) d\hat t.
\ee
(Here we choose units with $M$ the present Planck mass such that $dt'$ equals $dt$ at the present time.) For solutions with an exponential expansion, $\chi=\chi_0\exp(c\mu t)$, one finds
\be\label{E2}
t'=\frac{\chi_0}{c\mu M}\exp (c\mu t).
\ee
We observe that $t'$ goes to zero as $t\to-\infty$. In the new time coordinate $t'$ the big bang appears as the ``origin of time''. We emphasize that this is a pure coordinate effect. Expressing the solution \eqref{5} in the $t'$-coordinate the field $\chi$ increases linearly
\be\label{E2A}
\chi(t')=c\mu Mt'.
\ee

We may perform a similar rescaling of the space coordinates. It is more convenient, however, to use again fixed (comoving) space positions and to employ instead in the Robertson Walker metric a rescaled scale factor
\be\label{E3}
a'(t')=\frac{\chi(t)}{M}a(t).
\ee
Evaluating the Hubble parameter in the rescaled coordinates yields 
\ba\label{E4}
H'&=&\frac{d}{dt'}\ln a'=\frac{M}{\chi}\frac{d}{dt}(\ln a+\ln \chi)\nn\\
&=&\frac M\chi \left(H+\frac{\dot{\chi}}{\chi}\right).
\ea
The rescaled Hubble parameter $H'$ differs from $H$ in two important aspects. First, for constant $H=b\mu$ and $\dot{\chi}/\chi=c\mu$ one finds $H'$ to be proportional to $1/t'$,
\be\label{E5}
H'=\frac{b+c}{ct'}.
\ee
Second, even for a shrinking Universe with $b<0$ the Universe appears expanding in the new coordinates if $c>-b$. In particular, for the radiation and matter dominated epochs with $b=-c/2$ \eqref{B10} or $b=-c/3$ \eqref{B5} one finds an expansion $H'=1(2t')$ or $H'=2/(3t')$, respectively. This results in the familiar expansion laws $a'\sim (t')^{1/2}$ or $a'\sim(t')^{\frac32}$. It is in the rescaled coordinates that cosmology takes the usual form!

The transformations \eqref{E1}, \eqref{E3} are instructive  by emphasizing the important role of the units in which we measure time and space. We note, however, that they depend on the field $\chi(t)$ and are therefore only defined if some particular solution $\chi(t)$ is used. A more appropriate procedure uses field rescalings on the level of the effective action. This permits to discuss all possible solutions at once.

\medskip\noindent
{\bf {\em 2. Einstein frame.}}

Instead of rescaling the coordinates we may also keep a fixed coordinate system and change the metric. This amounts to a nonlinear transformation in the space of field variables which will change the form of the effective action and the field equations. The choice of field variables does not matter for physical observables which always concern dimensionless numbers, as ratios of masses. This has been demonstrated explicitly in ref. \cite{CW1,CW3}, where it has been argued that once the quantum effective action is computed (or assumed) both frames are equivalent in all respects, see also ref. \cite{DamE}. Explicit computations in the Jordan frame strengthen this point \cite{Fla,Cat1,Cat2,DeS,Ja}. (For a different point of view and an extensive, still incomplete in important aspects, documentation of the debate around this issue see ref. \cite{Far}.) Based on this equivalence many practical computations of observables are most easily done in the Einstein frame, and we will do this later for our discussion of 
cosmon inflation. Since Weyl scaling can be considered as a coordinate change in field space we should not be surprised 
that singularities may appear in certain field-coordinate systems. We will see that the big bang singularity in the Einstein frame is of this type. 

A Weyl scaling defines new metric variables $g'\mn$ by
\be\label{W1}
g\mn=\frac{M^2}{F(\chi)}g'\mn.
\ee
In these variables the action \eqref{1} reads
\ba\label{W2}
\Gamma=\int d^4x\sqrt{g'}\Big\{
&&\hspace{-0.4cm}-\frac{M^2}{2}R'+
\frac{M^2 K'}{2\chi^2}
\partial_\mu\chi\partial_\nu\chi g^{\prime\mu\nu}\nn\\
&&\hspace{-0.4cm}+V'(\chi)\Big\},
\ea
with $R'$ the curvature scalar associated to the metric $g'_{\mu\nu}$. The potential is rescaled according to
\be\label{W3}
V'=\frac{M^4 V}{F^2},
\ee
and the kinetial reads in the Einstein frame
\be\label{W4}
K'=\chi^2\left\{\frac KF+\frac32
\left(\frac{\partial \ln F}{\partial\chi}\right)^2\right\}.
\ee
Particle masses that scale $\sim \sqrt{F}$ in the ``Jordan frame'' (using $g\mn$) are constant in the ``Einstein frame'' (using $g'\mn$). In the following we will omit the prime on the metric. 

For the particular choice of $F$ in eq. \eqref{A1} one finds 
\ba\label{W5}
V'&=&\frac{M^4\mu^2}{\chi^2},\nn\\
K'&=&K+6.
\ea
Now the potential decreases for $\chi\to\infty$, in contrast to eq. \eqref{A1}. In fact, the only physical observables are dimensionless quantities. For the potential this is the ratio of $V$ divided by the fourth power of the effective Planck mass. This yields the effective cosmological constant in units of the Planck mass. In the Jordan frame \eqref{1}, \eqref{A1} the ratio reads $\mu^2\chi^2/\chi^4=\mu^2/\chi^2$, and the same value obtains in the Einstein frame \eqref{W2}, \eqref{W5}, $V'/M^4=\mu^2/\chi^2$. In the Jordan frame the ratio vanishes asymptotically for $t\to\infty$ despite the fact that $V$ increases with increasing $\chi$: the fourth power of the effective Planck mass $\chi^4$ simply increases faster. This explains why the effective cosmological constant vanishes asymptotically in our model. In fact, it has been observed long ago that models where in the Jordan frame the potential $V(\chi)$ increases for large $\chi$ less fast than $\chi^4$ solve the cosmological constant problem 
asymptotically \cite{CW1}. This setting was the reason for the first proposal of dynamical dark energy or quintessence \cite{CW3}. In the Einstein frame the asymptotic vanishing of the cosmological constant is reflected by the vanishing of $V'$ for $\chi\to\infty$, i.e. by the absence of a nonzero constant for $V'(\chi\to\infty)$. The naturalness of this scenario in the presence of quantum and thermal fluctuations has been discussed extensively in ref. \cite{CWEXP}.

The discussion of stability of our model is also particularly transparent in the Einstein frame. The condition $K>-6$ simply translates to $K'>0$, which provides for the correct sign of the scalar kinetic term. We can bring the scalar field closer to a canonical form by using for the scalar field the variable
\be\label{12A}
\varphi=\frac{2M}{\alpha}\ln \left(\frac\chi\mu\right).
\ee
Then the action
 \eqref{1} reads 
\ba\label{68A}
&&\Gamma=\int d^4x\sqrt{g}\left\{-\frac{M^2}{2}R\right.\nn\\
&&\quad \left.+\frac{k^2}{2}\partial^\mu\varphi\partial_\mu\varphi+M^4\exp \left(-\frac{\alpha\varphi}{M}\right)\right\},
\ea
with 
\be\label{A2}
k^2=\frac{\alpha^2(K+6)}{4}.
\ee
In the limit of constant $k^2$ the exponential potential in eq. \eqref{68A} provides the prototype for scaling solutions for dynamical Dark Energy \cite{CW3,CW2,RP,QU4,FJ,LAC}.

We may now concentrate on the particular kinetial \eqref{2}. For large $\chi^2/m^2$ the function $k^2$ becomes very close to one such that the scalar ``cosmon'' field has a kinetic term with standard normalization. The effective action \eqref{A2} describes a standard model for quintessence with an exponential potential. One recovers the known scaling solutions \cite{CW3}
for the radiation $(n=4)$ and matter $(n=3)$ dominated epochs, with a constant fraction of Early Dark Energy (EDE)
\be\label{14}
\Omega_e=\frac{n}{\alpha^2}.
\ee
This is precisely  the amount of EDE found in the Jordan frame, c.f. eqs. \eqref{B4}, \eqref{38A}.
One may verify that the de Sitter solutions \eqref{B11}, \eqref{B3} and \eqref{B6}, \eqref{B7} are in one to one correspondence with the standard scaling solutions. Observational bounds set typical limits $\Omega_e\lesssim 0.01$ \cite{A2a,A2b,Re,A2c,A2d,PL}, and we will adopt here a conservative bound $\alpha^2>100$. 

On the other hand, the very early stages of inflation with small $\tilde \alpha^2$ and therefore large $k^2\approx \alpha^2/\tilde \alpha^2$ correspond to power law inflation. We can rescale the field $\varphi$ in order to work with a standard kinetic term, which amounts to replace $\alpha$ by $\tilde \alpha$ in the exponential potential. We will see below that the phase of inflation which determines the properties of density fluctuations differs from standard power law inflation.

At this point it may be worthwhile to discuss the origin of the apparent singularity of the big bang in the Einstein frame for the metric. For the solutions of model (A)
we can associate the ``big bang'' with $\chi\to 0$. In the coordinate $t$ this happens for $t\to-\infty$, while for the coordinate $t'$ the big bang occurs for $t'\to 0$. 
The curvature tensor formed from the metric $g\mn$ remains finite, while it becomes singular for the metric $g'\mn$ at the time when $\chi$ reaches zero. 
The reason is simply that $R'$ is related to $R$ by a multiplicative factor $M^2/\chi^2$ which diverges for $\chi\to 0$. 
(The precise relation contains also additive terms involving derivatives of $\chi$.) 
We conclude that the usual ``big bang singularity'' is a ``coordinate effect'' in the space of field variables. 
There exist simple choices of fields where solutions are regular for all time.
In particular, we will discuss below a different frame where the geometry becomes flat Minkowski space with a constant scale factor as $t\to-\infty$.

We emphasize that the big bang singularity in the Einstein frame is not a coordinate singularity. Coordinate transformations leave invariants as $R$ or $R_{\mu\nu\lambda\sigma}R^{\mu\nu\lambda\sigma}$ unchanged and can therefore not remove singularities in these quantities.
Field transformations are a much wider concept than coordinate transformations. They can indeed remove singularities in quantities as $R$ as we just have demonstrated. We argue that a singularity  which disappears for an appropriate choice  of fields should not be considered as a  ``true'' or ``physical singularity''. It is rather an artefact of a choice of field-coordinates that becomes inappropriate under certain conditions.  We also note that  the issue of a big bang singularity is not related to inflation. For constant $K$ we have found a de Sitter solution for the radiation dominated Universe \eqref{B11}, \eqref{B3}, \eqref{B3a} which is regular for all $t$.

\medskip\noindent
{\bf {\em 3. Flat geometry for radiation or matter

~domination.}}

For the radiation and matter dominated epoch we have replaced the expanding Universe with constant Planck mass by a shrinking Universe with increasing Planck mass. We can also find coordinates where the Universe is static, while the Planck mass grows with a different rate. For this purpose we choose a more general coordinate $t'$ and rescaled scale factor $a'$ according to 
\be\label{G1}
dt'=\left(\frac\chi M\right)^\eta dt~,~a'=\left(\frac\chi M\right)^\eta a,
\ee
(with eqs. \eqref{E1}, \eqref{E3} realized for $\eta=1$.) This yields a rescaled Hubble parameter
\be\label{G2}
H'=\left(\frac M\chi\right)^\eta\left(H+\eta\frac{\dot{\chi}}{\chi}\right).
\ee
Solving eq. \eqref{G1} for the scaling solution \eqref{5} one finds 
\be\label{G3}
H'=\frac{1}{t'}\left(1+\frac{b}{\eta c}\right).
\ee
Choosing
\be\label{G4}
\eta=-\frac bc
\ee
the Hubble parameter vanishes, leading to a static  scale factor. For the matter and radiation dominated epochs the choice of coordinates for which the Universe is static corresponds to $\eta=1/3$ or $\eta=1/2$, respectively. For this choice the geometry is flat Minkowski space. 

This observation guides us to a use of field variables for which either the  radiation or matter dominated epoch is realized with flat Minkowski space. We can  transform the action \eqref{1}, \eqref{A1}
 by a different choice for the metric
\be\label{G5}
g_{\mu\nu}=w^2 g'_{\mu\nu}~,~w^2=\left(\frac{M^2}{\chi^2}\right)^\eta,
\ee
resulting in 
\be\label{G6}
R=w^{-2}\big[R'-6g'^{\mu\nu}(D_\mu D_\nu\ln w+\partial_\mu\ln w\partial_\nu\ln~ w)\big].
\ee
For $F=\chi^2$ the Weyl scaling \eqref{W1} corresponds to $\eta=1$.
  With $\varphi$ defined by eq. \eqref{12A} and omitting the primes the effective action reads in the new frame
\ba\label{G7}
\Gamma&=&\int d^4 x\sqrt{g}
\left\{\left(\frac{M^2}{\mu^2}\right)^{\eta-1}\exp\left(\frac{(1-\eta)\alpha\varphi}{M}\right)
\left[-\frac{M^2}{2}R\right.\right.\nn\\
&&\left.+\frac{\alpha^2}{8}(K+12\eta-6\eta^2)\partial^\mu\varphi\partial_\mu\varphi\right]\nn\\
&&\left.+    \mu^4 \left(\frac{M^2}{\mu^2}\right)^{2\eta}    \exp \left(\frac{(1-2\eta)\alpha\varphi}{M}\right)\right\}.
\ea
Different choices of $\eta$ define different choices of the  metric or different ``frames''.

For $\eta=1/3$ or $\eta=1/2$ the Hubble parameter vanishes in the matter or radiation dominated epoch,  respectively. Using the general formalism  of sect. II on infers from eq. \eqref{6A} the condition for a vanishing Hubble parameter,
\be\label{109A}
V+\frac{1}{2}\tilde K\dot\varphi^2+\rho=0,
\ee
with $(\chi^2\gg m^2)$
\ba\label{109B}
\tilde K&=& \frac{\alpha^2\mu}{4M}\exp\left(\frac{\alpha\varphi}{2M}\right)(K+12\eta-6\eta^2)\nn\\
&=&\frac{\mu}{M}\exp\left(\frac{\alpha\varphi}{2M}\right)\left[1-\frac{3\alpha^2}{2}(1-\eta)^2\right].
\ea
Since $\tilde K$ is negative both for the matter and radiation dominated phase  (recall $\alpha^2 > 3$ or $\alpha^2 > 4$, respectively) the negative  contribution $\sim \dot\varphi^2$ can cancel the positive contribution $V+\rho$.

It is instructive to consider the radiation dominated epoch in a frame  with $\eta=1/2$. The potential  $V=\mu^2 M^2$ is constant, and for a vanishing  Hubble parameter  also the radiation  density $\rho$ is constant. Eq. \eqref{109A} yields
\ba\label{109C}
\dot\varphi^2&=&-\frac{2(V+\rho)}{\tilde K},\nn\\
\tilde K&=&-\left(\frac{3\alpha^2}{8}-1\right)\frac{\mu}{M}\exp\left(\frac{\alpha\varphi}{2M}\right).
\ea 
Renormalizing  the scalar field,
\be\label{109D}
\sigma=\sqrt{M\mu}\exp \left(\frac{\alpha\varphi}{4M}\right),
\ee
eq. \eqref{109C} is easily solved,
\be\label{109E}
\sigma=\sqrt{\frac{\alpha^2 f_\rho}{3\alpha^2-8} }M\mu t\ ,\
f_\rho=1+\frac{\rho}{V}.
\ee
The ratio $f_\rho=3\alpha^2/4-2$ is determined by the solution of the scalar field equation, such that
\be\label{109E2}
\sigma = \frac{\alpha M \mu t}{2}.
\ee

In terms of $\sigma$ the effective action \eqref{G7} takes a simple form
\be\label{109F}
\Gamma=\int d^4 x\sqrt{g}\left\{-\frac{1}{2}\sigma^2 R+\mu^2 M^2+\frac{1}{2} K_\sigma\partial^\mu \sigma\partial_\mu\sigma\right\},\ee
with
\be\label{109G}
K_\sigma=-6+\frac{16}{\alpha^2}.\ee
We conclude that it has the same form as eq. \eqref{1}, except for a replacement of the potential  $\mu^2\chi^2$ by a constant $\mu^2 M^2$,  accompanied by a shift in the constant  kinetial from $-6+\frac{4}{\alpha^2}$ to $-6+\frac{16}{\alpha^2}$.
(The possibility of this interesting rescaling of the potential has already been noted in the appendix of ref. \cite{CW1}.) We observe that for large $\chi^2\gg m^2$ the effective action \eqref{109F} actually corresponds  to our model (B), eq. \eqref{2A}. The radiation dominated epoch of this model will be discussed in sect. IX. While the geometry is static flat space, all particle masses increase proportional to the effective Planck mass $\sigma$. This is how the predictions of nucleosynthesis are reproduced to be the same as in standard cosmology.

For a frame with $\eta=1/3$ the geometry turns out to be Minkowski space for the matter dominated period. Now $\rho$ increases proportional to the particle masses, that are in turn proportional to the effective Planck mass $\sim\exp(\alpha\varphi/3M)$. This permits to realize a matter dominated Universe without expansion, as pursued in the very early days of cosmology based on general relativity.  The prize to pay is, however, a non-static behavior of the particle masses.
We observe that for the frame with $\eta=1/3$ the future scalar dominated Universe is expanding, according to the positive sign of $b+\eta c= 2\tilde\gamma c/3$, cf. eq. \eqref{N9}.

For $\eta\not= 0,1$ the choice of the metric corresponds neither to the Einstein frame  (fixed Planck mass) nor to the Jordan frame (scale transformations realized by multiplicative scaling of fields). A ``flat frame'' may be defined by a choice of fields for which the geometry of the Universe is flat and static Minkowski space. Thus the choice  $\eta=\frac12$ constitutes a flat frame for the radiation dominated epoch, while $\eta=\frac13$ corresponds to the flat frame for the matter dominated epoch. We observe that no choice of $\eta$ provides for a flat frame for both the radiation {\em and} matter dominated epochs. (Formally, it may be possible to construct a frame with flat space for the whole period of radiation and matter domination, using a rather complicated function $w(\chi/M)$ in the Weyl scaling \eqref{G5}. This function would have to mimic the details of the transition from radiation to matter domination, involving in turn detailed particle masses and interactions. Such a formulation looks not 
very natural.)

A flat frame for the matter dominated epoch has been proposed earlier in ref. \cite{Na1,Na2,Na3}. Our model differs from this proposal by the form of the cosmon coupling to the matter fields which involves the different particle masses according to eq. \eqref{8Y} or \eqref{8Z}, and by the different kinetic term for the cosmon.
This allows us to obtain a realistic description of nucleosynthesis or the cosmic microwave background, in contrast to the earlier proposals \cite{Na1,Na2,Na3}. Furthermore, the presence of a cosmon potential $V(\chi)$ plays an important role in our model. It dominates the cosmic evolution during inflation and in the present dark energy dominated epoch.

\medskip\noindent
{\bf {\em 4. Initial state with flat geometry.}}

It is also possible to describe the scaling solutions describing the inflationary epoch in a frame where geometry becomes flat Minkowski space in the limit $\chi\to 0$. In a somewhat different context this property has been observed in ref. \cite{Pi} for other models of inflation.  

In our context we choose
\be\label{109H}
\eta=-\frac{K+4}{2}=1-\frac{2}{\tilde\alpha^2}.\ee
Defining
\be\label{109I}
\sigma=\left(\frac{\mu}{M}\right)^{\frac{1}{2\tilde\alpha^2}}M\exp\left(\frac{\alpha\varphi}{\tilde\alpha^2 M}\right)\ee
the effective action reads
\be\label{109J}
\Gamma=\int d^4 x\sqrt{g}\left\{-\frac{\sigma^2}{2}R+\bar\mu^{\tilde\alpha^2}\sigma^{4-\tilde\alpha^2}+\frac{1}{2} K_\sigma\partial^\mu\sigma\partial_\mu\sigma\right\},\ee
where
\be\label{109K}
K_\sigma=\tilde\alpha^2-6\ee
and
\be\label{109L}
\bar\mu=\left(\frac{\mu}{M}\right)^{\frac{6}{\tilde\alpha^4}+\frac{1}{2\tilde\alpha^2}}M.\ee 
With these field variables the intrinsic scale $\bar\mu$ is tiny as compared to $\mu$ if $\tilde\alpha$ is small. Comparing with the equivalent  effective action \eqref{1}, \eqref{A1} this demonstrates that the role of the intrinsic scale strongly depends on the
choice of  coordinates in field space. 

For the initial  scaling solution at the beginning of inflation one has
\be\label{109M}
V+\frac{1}{2} K_\sigma\dot\sigma^2=0,\ee
such that $H=0$ according to eq. \eqref{6A}. The solution of eq. \eqref{109M},
\ba\label{109N}
\sigma&=&\sigma_0\left(1-\frac{\bar\mu t}{\kappa}\right)^{-\left(\frac{1}{1-\tilde\alpha^2/2}\right)},\nn\\
{\kappa}&=&\frac{\sqrt{3-\tilde\alpha^2/2}}{1-\tilde\alpha^2/2}\left(\frac{\bar\mu}{\sigma_0}\right)^{1-\frac{\tilde\alpha^2}{2}},\ea
remains regular for $t\to -\infty$ where $\sigma$ goes to zero. On the other hand, $\sigma$
becomes singular for $\bar\mu t\to{\kappa}$. This ``future singularity'' occurs, however,  far outside the region of interest. (We know already that it is absent for other choices of field coordinates.) For the inflationary epoch one has $\bar\mu t/{\kappa}\ll 1$ and therefore to a good approximation the linear increase
\be\label{109P}
\sigma=\sigma_0+\frac{\sigma^2_0}{\sqrt{3-\frac{\tilde\alpha^2}{2}}}\left(\frac{\bar\mu}{\sigma_0}\right)^{\frac{\tilde\alpha^2}{2}}t.\ee

The ``initial period'' of the cosmological  history is strikingly simple in the field coordinates \eqref{109J}. 
It is described by a slow  increase of a scalar field in flat Minkowski space. 
The potential realizing this picture is almost the scale invariant $\sigma^4$-potential, up to a small anomalous dimension $\sim \tilde\alpha^2$. 
It has a minimum at $\sigma=0$.
The field equations have a particular static solution
\be\label{127AX}
\sigma = 0, ~ H=0, ~ a=const.
\ee
This solution will be approached for $t \to -\infty$.
Thus what is called usually the ``big bang'' becomes in this picture Minkowski space with a static vanishing value of the scalar field.

The solution \eqref{127AX} is unstable for increasing $t$ as given by eq. \eqref{109N} for arbitrarily small nonzero $\sigma_0$.
For small initial $\sigma_0$ the field $\sigma(t)$ remains almost constant. 
Nevertheless, for $\sigma_0>0$ the small gradient $\partial V/\partial\sigma$ generates a slow time evolution, and the field increases due to the negative  value of $K_\sigma$. 
In flat space and for constant $K_\sigma$ the scalar field equation,
\be\label{109Q}
K_\sigma\ddot\sigma=-\frac{\partial V}{\partial\sigma},\ee
implies the conservation of 
\be\label{109R}
E_\sigma=V+K_\sigma \dot\sigma^2 /2,\ee
and the particular solution consistent with flat space according to eq. \eqref{109M} is the one for $E_\sigma=0$. We have seen in sect V that this solution is stable and attractive as  time increases. On the other hand, solutions  of eq. \eqref{109Q} with $E_\sigma\not= 0$ do not solve the combined system of field equations due to the gravitational coupling $\sigma^2 R$.

In this context we may work out the general condition for the existence of a solution with flat space in the framework of the effective action \eqref{1} in the absence of matter and radiation. The field equations \eqref{3}-\eqref{6A} have to be obeyed for $R=0,~H=0,~T_{\mu\nu}=0,~q_\chi=0$. Eq. \eqref{6} reads
\be\label{131A1}
K\ddot{\chi}=-\frac{\partial V}{\partial \chi}-\frac12\frac{\partial K}{\partial\chi}\dot{\chi}^2,
\ee
implying that $V+\frac12 K\dot{\chi}^2$ is conserved. For $K\neq 0$ eq. \eqref{6A} becomes
\be\label{131A2}
\dot{\chi}^2=-\frac{2V}{K}.
\ee
We consider the particular case $\partial F/\partial \chi^2=1$ such that insertion of eqs. \eqref{131A1}, \eqref{131A2} into eq. \eqref{6A} yields the condition
\be\label{131A3}
\frac{\partial\ln V}{\partial \ln \chi}-\frac{\partial\ln K}{\partial \ln \chi}=-(K+2).
\ee
This is in accordance with eqs. \eqref{109J}, \eqref{109K}, with $\chi$ replaced by $\sigma$.

\medskip\noindent
{\bf {\em 5. Future flat space.}}

The future cosmology of our model is dominated by the cosmon coupled to neutrinos. The choice of a frame for which the Universe approaches flat space in the future  therefore depends on the detailed $\chi$-dependence of the neutrino mass. We report here only briefly the approximation of constant $\tilde\gamma$ according to eq. \eqref{N6}. In this case we can employ again a frame  \eqref{G5} with constant $\eta$. The condition \eqref{109A} for a vanishing Hubble parameter requires for the flat frame that
$V$ and $\rho$ scale with the same power of $\chi$. For $m_\nu\sim\chi^{2\tilde\gamma+1}$ in the Jordan frame one has
$m'_\nu=w m_\nu\sim\chi^{2\tilde\gamma+1-\eta}$ in the frame with constant $\eta$. In this frame the potential scales as $V'=V w^4\sim\chi^{2-4\eta}$, such that the required scaling property is realized for
\be\label{126A}
\eta=\frac{1-2\tilde\gamma}{3}.\ee
According to eq. \eqref{N9} this reproduces $\eta=-b/c$. (For $\tilde\gamma=0$ one recovers the flat frame for matter domination.)

As a consequence,  the flat frame is characterized by the scaling 
\be\label{126B}
m'_\nu\sim V'\sim\chi^{\frac23(1+4\tilde\gamma)}.\ee
In this ``future flat frame'' the masses of the charged particles increase as well
\be\label{126C}
m'_c\sim m_c w\sim\chi^{1-\eta}\sim\chi^{\frac23(1+2\tilde\gamma)}.\ee
Due to the slower increase as compared to the neutrino mass their role becomes negligible in the future. In the future flat frame \eqref{126A} the Universe is shrinking during matter domination, according to the negative sign of $b+\eta c=-2\tilde\gamma c/3$.

\medskip\noindent
{\bf {\em 6. Present Universe in different frames.}}

The present  epoch is characterized by a transition from the  matter dominated epoch to a future dark energy dominated epoch for which the charged particles become irrelevant. Due to this transition character the geometry is, in general, non-static for all the simple frames with constant $\eta$. It may be instructive to describe the behavior in some of the different frames discussed above.

In the Einstein frame $(\eta=1)$ we have the accelerated expansion of standard cosmology in the presence of dark energy. In contrast, for the Jordan frame $(\eta=0)$ the present epoch witnesses  a transition from a shrinking Universe during matter domination to an expanding Universe in the future. This implies that there is a turning point for the scale factor at some given moment of the present cosmological era. For the flat frame for matter domination $(\eta=\frac13)$ the Universe has started to expand only recently, while for redshift $z\gtrsim 10$ it has
been in the static state for matter domination. Finally, for the future flat frame $(\eta=(1-2\tilde\gamma)/3)$ the Universe makes a transition from a shrinking scale factor during matter domination to a static state in the future.

A turning point from a shrinking to an expanding Universe exists for a large range of frames with
\be\label{126D}\frac{1-2\tilde\gamma}{3}<\eta<\frac13.\ee
The precise location of the turning point depends on the choice of $\eta$. For $\eta\to 1/3$ this point moves to the far past, and for $\eta\to(1-2\tilde\gamma)/3$ it occurs in the far future. By continuity there exists a value $\eta_t$ for which the Universe is static just at the present time. In this particular frame  the scale factor shrinks in the past and expands in the future. Just at the present time the Universe is static and the usual redshift of not too distant objects is purely due to the change of the mass of particles.

\section{Cosmon inflation for quadratic
cosmon potential}
\label{Cosmon inflation}

The early inflationary phase of the cosmology of our model is most easily described in the Einstein frame.
The effective action \eqref{68A} describes a standard  theory of gravity coupled to a scalar field with an exponential potential. The $\varphi$-dependence of the kinetial $k^2(\varphi)$ will, however, lead to quantitative differences from power law inflation.

\medskip\noindent
{\bf {\em 1. Kinetial and slow roll parameters.}}

The exponential form of the potential makes the slow roll formalism particularly simple. We will use in the next section a more general form of the potential in the Jordan frame that can be brought to a standard exponential form
in the Einstein frame by a suitable choice of $\varphi$. This will modify the particular form of the kinetial $k^2(\varphi)$, and we keep our formulae therefore general for arbitrary positive $k^2(\varphi)$. Our models are particular cases of ``cosmon inflation'' as it has been discussed recently in a wider context \cite{CI}. For the particular model \eqref{A1}, \eqref{2} the kinetial reads 
\be\label{C1}
k^2(\varphi)=\left(\frac{\alpha^2}{\tilde \alpha^2}-1\right)
\frac{m^2}{m^2+\mu^2\exp (\alpha\varphi/M)}+1.
\ee
It only varies substantially once $\chi^2=\mu^2\exp(\alpha\varphi/M)$ reaches values of the order $m^2$.

The properties of primordial density fluctuations with a given scale are governed by the properties of the potential and kinetial at the value of $\varphi$ at which the corresponding scale has left the horizon. The Hubble parameter during the slow roll phase can be approximated by 
\be\label{C6}
H^2=\frac{V}{3M^2}=\frac{M^2}{3}\exp \left(-\frac{\alpha\varphi}{M}\right).
\ee
We will see that important properties of the density fluctuations, as the spectral index $n$ and the tensor to scalar ratio $r$ depend only on the kinetial $k^2$. 
The overall amplitude of the fluctuations involves, in addition, the overall magnitude of $V$ at horizon crossing as given by the corresponding value of $\varphi$.
We present here only the most important features and refer for more details to ref. \cite{CI}.

The scalar field $\sigma$ with canonical normalization of the kinetic term is related to $\varphi$ by 
\be\label{C2}
\frac{d\sigma}{d\varphi}=k(\varphi).
\ee
We can use this relation in order to make direct contact with the usual treatment of single field inflation. In particular, it is straightforward to compute the standard slow roll parameters of inflation as a function of $\varphi$
\ba\label{C3}
\epsilon&=&\frac{M^2}{2}\left(\frac{\partial\ln V}{\partial\sigma}\right)^2=\frac{M^2}{2k^2}
\left(\frac{\partial\ln V}{\partial\varphi}\right)^2=\frac{\alpha^2}{2k^2},\nn\\
\eta&=&\frac{M^2}{V}\frac{\partial^2V}{\partial\sigma^2}=2\epsilon-\frac{M}{\alpha}\frac{\partial\epsilon}{\partial\varphi}.
\ea
They depend only on the kinetial $k^2(\varphi)$.

The number of $e$-foldings before the end of inflation obeys the simple relation 
\be\label{53A}
N(\varphi)=\frac{1}{\alpha M}\int\limits^{\varphi_f}_\varphi d\varphi'k^2(\varphi').
\ee
Inverting this relation and inserting  into eq. \eqref{C3} yields $\epsilon(N)$ and $\eta(N)$.
The formulae \eqref{C3}, \eqref{53A}
are valid for an arbitrary form of the kinetial $k^2(\varphi)$. Only the kinetial $k^2(\varphi)$ enters in the computation of $\epsilon (N),\eta (N)$ and $N$. In turn, the spectral index $n$ and the tensor to scalar ratio $r$ for the density perturbations generated during inflation are given by 
\ba\label{53B}
n&=&1-6\epsilon(N)+2\eta(N),\nn\\
r&=&16\epsilon(N),
\ea
where $N\approx 50-60$, depending on details of the entropy production after the end of inflation. Thus $n$ and $r$ can be determined for any given form of $k(\varphi)$. 
We observe  that by a multiplicative rescaling of $\varphi$ in eq. \eqref{68A} one can obtain $\alpha=1$. Thus   $\epsilon,\eta$ and $N$ involve only  the combination  $k^2/\alpha^2$.

\medskip\noindent
{\bf {\em 2. Slow roll parameters for quadratic cosmon 

~potential.}}

In our context, these quantities depend  on the two parameters $\tilde \alpha$ and $\alpha$. For the specific form of the kinetial \eqref{C1} the integral \eqref{53A} can be solved explicitly, 
\ba\label{53C}
N(\varphi)
&&=\frac{\alpha(\varphi_f-\varphi)}{\tilde \alpha^2M}\\
&&-\left(\frac{1}{\tilde \alpha^2}-\frac{1}{\alpha^2}\right)\ln\left(\frac{m^2+\mu^2\exp(\alpha\varphi_f/M)}{m^2+\mu^2\exp(\alpha\varphi/M)}\right).\nn
\ea
This makes a numerical computation of $n(N)$ and $r(N)$ very easy and we will present results below.  The qualtitative features can be understood by simple analytic considerations.

An inflationary phase requires a range of $\varphi$ for which $\epsilon\ll 1,|\eta|\ll 1$. This can be realized for small or negative $\varphi$ provided $\tilde \alpha\ll 1$. 
Then the $\varphi$-dependence of $k^2$ can be neglected such that $\epsilon=\tilde \alpha^2/2,\eta=\tilde \alpha^2$. 
On the other hand, for large $\varphi$ the kinetial approaches $k^2=1$. 
For the large values of $\alpha^2>100$ needed in order to obey the restrictions on early dark energy this implies that the slow roll period has to end for a sufficiently large value of $\varphi$. 
For large $\alpha\gg 1$ and small $\tilde \alpha\ll 1$ we can approximate the slow roll period and its end by
\ba\label{C4}
\epsilon&=&\frac{\tilde\alpha^2}{2}\left(1+\frac{\mu^2}{m^2}\exp (\alpha\varphi/M)\right),\nn\\
\eta&=&\epsilon+\frac{\tilde \alpha^2}{2}.
\ea
The inflationary phase ends when $\epsilon$ reaches one, corresponding to a value $\varphi_f$ obeying
\be\label{C5}
\exp \left(\frac{\alpha\varphi_f}{M}\right)=\frac{2m^2}{\tilde \alpha^2\mu^2}.
\ee
Inserting this value in eq. \eqref{53C} and neglecting $\alpha^{-2}$ as compared to $\tilde\alpha^{-2}$ yields
\ba\label{57A}
N(\varphi)\approx \frac{1}{\tilde \alpha^2}&&\hspace{-0.3cm}\left[\ln\left(\frac{m^2}{\mu^2}+\exp\left(\frac{\alpha\varphi}{M}\right)\right)
-\frac{\alpha\varphi}{M}\right. \nn\\
&&\hspace{-0.3cm}\left.-\ln\left(1+\frac{\tilde \alpha^2}{2}\right) \right].
\ea

At this point we can extract the value of $\varphi$ for a given number of $e$-foldings before the end of inflation and determine $\epsilon, \eta,n,r$ according to eqs. \eqref{C4}, \eqref{53B}. Since $\alpha$ appears only in the combination $\alpha\varphi/M$ the result does not depend on $\alpha$. We display the values of $n$ and $r$ for different values of $\tilde \alpha$ in Table 1. The first number corresponds  to $N=60$, and the second number in brackets refers to $N=50$.

\begin{table}[h!]\label{Table1}
\centering
\begin{tabular}{|p{10mm}|p{20mm}|p{20mm}|p{20mm}|}\hline
\centering{$\tilde\alpha$}&\centering{$0.001$}&\centering{$0.02$}&\centering{$0.1$}\tabularnewline\hline
\centering{$n$}& \centering{0.975 (0.97)} 
&\centering{0.975 (0.97)} &\centering{0.972 (0.967)}\tabularnewline \hline
\centering{$r$}&\centering{0.13 (0.16)}&\centering{0.13 (0.16)}&\centering{0.18 (0.2)}\tabularnewline \hline
\centering{$\frac m\mu$}&\centering{120 (100)}&\centering{2400 (2000)}&\centering{12\,000(10\,000)}\tabularnewline\hline
\end{tabular}
\caption{Properties of density fluctuations, model (A).}
\end{table}

\medskip\noindent
{\bf {\em 3. Late horizon crossing.}}

We may distinguish two scenarios for the ratio 
\be\label{C8}
x=\left(\frac{\chi^2}{m^2}\right)=\left(\frac{\mu^2}{m^2}\right)\exp(\alpha\varphi/M)
\ee
at the time of horizon crossing of the fluctuations.  In terms of $x$ the kinetial reads
\be\label{102A}
k^2=\left(\frac{\alpha^2}{\tilde\alpha^2}-1\right)(1+x)^{-1}+1. 
\ee
For large $x\gg 1$ we can approximate the kinetial and slow roll parameters by 
\be\label{102B} 
k^2=\frac{\alpha^2}{\Aa x}~,~\epsilon=\frac12\eta=\frac12\Aa x.
\ee
They only depend on the parameter $\tilde \alpha$ or equivalently $K(\chi=0)$. No intrinsic mass scale appears for the slow roll parameters. Their determination needs for a given $\tilde \alpha$ the value $x(\tilde \alpha)$ for horizon crossing. In principle, the combination $\Aa x(\tilde\alpha)$ depends on $\tilde \alpha$. We find, however, that horizon crossing in the asymptotic regime requires small $\tilde \alpha$. Since $\tilde \alpha^2 x(\tilde \alpha)$ has a finite value for $\tilde \alpha\to 0$, this value will yield a ``universal value'' (independent of all parameters) for the properties of density fluctuations. 

In order to determine $x(\tilde \alpha)$ we use 
\be\label{57B}
N(\varphi)\approx \frac{m^2}{\tilde \alpha^2\mu^2}\exp\left(-\frac{\alpha\varphi}{M}\right)=\frac{1}{\tilde \alpha^2 x}.
\ee
In the large-$x$-region we therefore find the simple relations
\be\label{C9}
x=\frac{1}{N\tilde \alpha^2}~,~\epsilon=\eta=\frac{\tilde \alpha^2x}{2}=\frac{1}{2N}.
\ee
As a consequence, the spectral index $n$ and the tensor to scalar ratio of the primordial density fluctuations do not depend on the parameters $\alpha,\tilde \alpha,\mu^2/m^2$,
\ba\label{C10}
n&=&1-6\epsilon +2\eta=1-\frac2N,\nn\\
r&=&16\epsilon=\frac8N=4(1-n).
\ea
For $N=60$ eq. \eqref{C10} implies
\be\label{C9a}
n\approx 0.97~,~r\approx 0.13.
\ee
Comparison with Table I shows that for $\tilde\alpha=0.001$ or $\tilde\alpha=0.02$ both  $n$ and $r$ are well approximated by eq. \eqref{C10}.

\medskip\noindent
{\bf {\em 4. Early horizon crossing.}}

The other limiting regime corresponds to $x\ll 1$, where one approximates
\be\label{58A}
N=\frac{1}{\tilde \alpha^2}\left(\ln\left(\frac1x\right)+x-\ln\left(1+\frac{\tilde \alpha^2}{2}\right)\right).
\ee
Concentrating on the leading term $\sim\ln(1/x)$ one finds
\be\label{58B}
x=\exp (-\Aa N)
\ee
and therefore, for $x\to 0$,
\be\label{58C}
\epsilon=\frac12\eta=\frac{\Aa}{2}.
\ee
The transition between the two limiting regions occurs for $\Aa\approx 1/N$, with $x\gg 1$ realized for $\Aa\ll 1/N$ and $x\ll 1$ for $\Aa\gg 1/N$. 
Thus $\epsilon$ tends to increase from the value \eqref{C10} as $\tilde \alpha$ increases. 
In view of the limits on $r$ extracted from the CMB anisotropies \cite{PL} a horizon crossing in the large $x$ region is favored.
It has to be investigated if the rather strong tensor amplitude $r\approx 0.13$ is compatible with observation once EDE and growing neutrino masses are included in the analysis of the CMB anisotropies.

\medskip\noindent
{\bf {\em 5. Amplitude of density fluctuations.}}

The dimensionless parameters $\tilde \alpha,\alpha$ and $\mu^2/m^2$ are further restricted by the requirement that the amplitude of primordial density fluctuations coincides with the observed one,
\be\label{C8a}
24\pi^2\Delta^2=\frac{V}{\epsilon M^4}=2N\exp \left(-\frac{\alpha\varphi}{M}\right)\approx 5\cdot 10^{-7}.
\ee
This requires  (for $N=60$)
\ba\label{62A}
\exp \left(-\frac{\alpha\varphi}{M}\right)=\frac{\mu^2}{m^2x}&\approx& 4\cdot 10^{-9},.
\ea
For $\Aa\ll 1/N$ this results in the condition
\ba\label{C9b}
\frac{\tilde \alpha^2\mu^2}{m^2}&\approx& \frac23\cdot 10^{-10}.
\ea
The values of $m^2/\mu^2$ for other values of $\tilde \alpha$ are also indicated in Table 1. Typical parameters for the realization of a realistic inflationary phase are $m\approx 100\mu,\tilde\alpha^2=(2/3)10^{-6}$, such that the value of $x$ relevant for the density fluctuations becomes $x=2.5\cdot 10^4$.

\medskip\noindent
{\bf {\em 6. More general kinetials.}} 

For the parameter region of very small $\tilde\alpha^2$ the function $K$ becomes very large for $\chi^2\to 0$. The behavior for $\chi^2\to 0$ obeys then the de Sitter solution \eqref{B6a},\eqref{B8a} with almost constant $\chi$,
\be\label{10a}
b\approx \frac{1}{\sqrt{3}}~,~c\approx 0~,~3H^2\approx \mu^2.
\ee
We emphasize that our setting holds for a wide variety of kinetials $\tilde k^2(x)=K(x)+6=4k^2(x)/\alpha^2=2/\epsilon(x)$, provided they have the property that they become very large for $x\to 0$ and small for $x\to\infty$. For example, one could replace eq. \eqref{2} by
\be\label{C11}
K+6=\frac{\hat m^2}{\chi^2}+\frac{g}{\ln(\chi^2/\hat m^2)}~,~\hat m^2=\frac{4m^2}{\tilde \alpha^2}.
\ee

\section{Flat cosmon potential with
Einstein term} 
\label{Flat cosmon potential}

So far we have concentrated on the simple class of models obeying eq. \eqref{A1}. 
The basic features of our setting hold for a much wider class of models of the type \eqref{1}. 
Whenever the potential $V(\chi)$ increases for large $\chi$ less fast than $F^2(\chi)$ the effective cosmological constant vanishes for asymptotic time as $\chi\to\infty$. 
For $F(\chi\to\infty)=\xi\chi^2$ and $K(\chi\to\infty)\to K_\infty$ the evolution for $\chi\to\infty$ corresponds to the approach to a fixed point with exact dilatation symmetry. 
Whenever the effective kinetic term obeys asymptotically $(K_\infty+6)/\xi\ll1$ the late cosmology obeys scaling solutions with a small fraction of early dark energy. 
On the other hand, an inflationary epoch occurs if $\chi^2 K/F$ is large for some range of $\chi$. 

In this section we discuss our second model (B), as defined by eqs. \eqref{1}, \eqref{2A}. 
The potential $V$ is now given by a cosmological constant $\bar\lambda_c$. 
Due to the increase of $F\sim\chi^2$ for large $\chi$ its dynamical role is very different from the one in Einstein gravity. 
We have already encountered  a model with constant $V$ in sect. VII, eqs. \eqref{109F}, \eqref{109G}. For $\chi^2\gg m^2$ model (B) will indeed coincide with eqs. \eqref{109F}, \eqref{109G}. 
For the radiation and matter dominated epochs there will be no difference between models (A) and (B), and this extends to the late dark energy dominated epoch. 

Besides the different potential a second important difference between the models (A) and (B) concerns the presence of an ``Einstein term'' $\sim -m^2 R$ in the effective action. 
The coefficient of the curvature scalar no longer vanishes for $\chi\to 0$. In short, the limiting behavior of the effective action for $\chi\to\infty$ realizes dilatation symmetry for both models (A) and (B). 
The predictions of both models are the same after the onset of radiation domination. On the other hand, 
the inflationary phase and its end is related to explicit scale symmetry breaking. Here both models differ.

\medskip\noindent
{\bf {\em 1. Effective action and field equations for flat 

~cosmon potential.}}

The effective action of model (B) involves
\be\label{SB1}
F(\chi)=\chi^2+m^2~,~V(\chi)=\bar\lambda_c.
\ee
The explicit scale symmetry breaking occurs by the presence of parameters with nonzero dimension. In the class of models \eqref{SB1} this concerns a cosmological constant (in the Jordan frame) $\bar\lambda_c$ and a violation of the scaling $F\sim\chi^2$ for small $\chi$ by a constant $m^2$. Since $F$ approaches now a constant for small $\chi$ the kinetial must be positive in this region, $K(\chi\to 0)\geq 0$. Realistic values for early dark energy require that $K(\chi\to\infty)$ is close to the critical value $-6$. Thus the form $F=m^2+\chi^2$ requires that the scale symmetry breaking is also present in the kinetial $K(\chi)$ in the form of a non-trivial $\chi$ dependence. We will again take $K(\chi)$ similar to eq. \eqref{2}. We will, however, choose a different normalization of $\alpha$ and $\tilde\alpha$ such that 
\be\label{SB1A}
K+6=\frac{16}{\tilde\alpha^2}\frac{m^2}{m^2+\chi^2}+\frac{16}{\alpha^2}\frac{\chi^2}{m^2+\chi^2}.
\ee
The contribution $\sim\tilde\alpha^{-2}$ ensures the positivity for $\chi\to 0$ provided $\Aa<8/3$. The inflationary phase of this model is again a special case of cosmon inflation, as discussed in ref. \cite{CI}.

Our model (B) has two characteristic mass scales, $(\bar\lambda_c)^{1/4}$ and $m$. We can associate $\bar\lambda_c$ with the present dark energy density, 
\be\label{SB2}
\frac{\bar\lambda_c}{M^4}\approx 7\cdot 10^{-121}~,~(\bar\lambda_c)^{1/4}=2\cdot 10^{-3} eV.
\ee
In the Jordan frame the potential energy for the scalar field remains the same at all time - it is the same during inflation and today. 
Only the ratio of $V$ to the fourth power of the effective Planck mass $F^{1/2}(\chi)$ changes as $\chi$ increases. In the Einstein frame this will lead again to an exponential potential for large $\varphi$.

We will find that the second mass scale $m$ has to be somewhat larger than $(\bar\lambda_c)^{1/4}$ in order to ensure the correct amplitude for the density perturbations generated during inflation. 
We will find below $m\approx 1$eV. The two mass scales $m$ and $(\bar\lambda_c)^{1/4}$ are the only intrinsic mass scales of our model. In addition, we have the mass scale generated by the spontaneous breaking of dilatation symmetry by a nonzero value of $\chi$. The masses of hadrons and charged leptons are supposed to scale proportional to $\chi$ for large $\chi$.

For the model \eqref{SB1} the field equations \eqref{A6}, \eqref{A7} read 
\ba\label{SB3}
&&\left(K+\frac{6\chi^2}{\chi^2+m^2}\right)
(\ddot{s}+3H\dot{s}+2\dot{s}^2)\nn\\
&&
+\left(\frac{\chi}{2}\frac{\partial K}{\partial\chi}-\frac{m^2 K}{\chi^2+m^2}\right)\dot{s}^2\nn\\
&&\qquad\quad  =\frac{4\bar\lambda_c}{\chi^2+m^2}+\frac{q_\chi}{\chi}-\frac{T^\mu_\mu}{\chi^2+m^2},
\ea
and 
\ba\label{SB4}
&&\left(H+\frac{\chi^2}{\chi^2+m^2}\dot{s}\right)^2\\
&&=\frac{1}{3(\chi^2+m^2)}\left[\bar\lambda_c+\frac{\chi^2}{2}\left(K+\frac{6\chi^2}{\chi^2+m^2}\right)\dot{s}^2+T_{00}\right].\nn
\ea
For our choice \eqref{SB1A} for $K$ the stability requirement
\be\label{SB5}
K>-\frac{6\chi^2}{\chi^2+m^2}
\ee
is obeyed for $\Aa<8/3$. 

We will take $\bar\lambda_c>0$. For $\rho=T_{00}>0$ the r.h.s. of eq. \eqref{SB4} is positive for all $s$ and $\dot{s}$ such that the sign of the combination 
\ba\label{SB6}
H+\frac{\chi^2}{\chi^2+m^2}\dot{s}&=&H+\frac{\chi\dot{\chi}}{\chi^2+m^2}\\
&=&\partial_t\big (\ln a+\frac12\ln(\chi^2+m^2)\big)\nn
\ea
cannot change during the time evolution. Similar to our model (A) the combination
\be\label{SB7}
y=\ln a+\frac12\ln\left(\frac{\chi^2+m^2}{m^2}\right)
\ee
is either monotonically increasing or decreasing. We choose time conventions such that $\dot{y}$ is positive, resulting in a positive root 
\be\label{107A}
H=-\frac{\chi^2}{\chi^2+m^2}\dot{s}+
\sqrt{\frac{\bar\lambda_c+T_{00}}{3(\chi^2+m^2)}+\frac{K'\dot{s}^2}{6}},\nn\\
\ee
with
\be\label{SB5a}
K'=\frac{\chi^2}{\chi^2+m^2}\left(K+\frac{6\chi^2}{\chi^2+m^2}\right).
\ee

\medskip\noindent
{\bf {\em 2. Radiation domination.}}

We are interested in simple scaling solutions and concentrate first on $\chi^2\gg m^2$. In this limit one finds a  negative constant (c.f. eq. \eqref{109G}) 
\be\label{SB6a}
K=-6+\frac{16}{\alpha^2}.
\ee
A simple scaling solution is found for 
\be\label{SB7a}
\rho=\bar\rho={\rm const}.~,~H=0.
\ee
In this case eq. \eqref{SB4} becomes 
\be\label{SB8}
\bar\lambda_c+\bar\rho+\frac{K}{2}\dot\chi^2=0,
\ee
with solution 
\be\label{SB9}
\chi=\sqrt{-\frac{2}{K}
(\bar\lambda_c+\bar\rho)}(t+t_0).
\ee
Inserting
\be\label{SB10}
\dot{s}^2=-\frac{2(\bar\lambda_c+\bar\rho)}{K\chi^2}\ ,\ 
\ddot{s}=-\dot{s}^2,
\ee
eq. \eqref{SB3} yields
\be\label{SB11}
-\frac{2(K+6)}{K}(\bar\lambda_c+\bar\rho) 
=4\bar\lambda_c+ q_\chi\chi-T^\mu_\mu.
\ee

Our assumption of constant $\rho$ has to be compatible with the conservation of the energy momentum tensor. For $H=0$ this holds for radiation, while matter with particle masses $\sim\chi$ would have $\rho$ changing $\sim\chi$. Our ansatz therefore describes the radiation dominated epoch, for which $q_\chi=0,T^\mu_\mu=0$. Eq. \eqref{SB11} turns to a simple algebraic equation fixing $\bar\rho/\bar\lambda_c$,
\be\label{SB12}
\frac{\bar\rho}{\bar\lambda_c}=-\frac{3(K+2)}{K+6}.
\ee
This is compatible with positive $\rho$ provided $(K>-6)$
\be\label{SB13}
K<-2.
\ee
For the asymptotic form of our choice for $K$ \eqref{SB1A} this requires, similar to our first model (A),
\be\label{SB14}
\alpha>2.
\ee

We conclude that the radiation dominated epoch becomes very simple for the model (B) \eqref{SB1}, \eqref{SB1A}. 
The static geometry is given by flat Minkowski space. 
The only time evolution concerns the linear increase of the effective Planck mass 
\be\label{SB15}
\chi=2\sqrt{\frac{\bar\lambda_c}{K+6}}(t+t_0) =\frac{\alpha}{2}\sqrt{\bar\lambda_c}(t-t_0).
\ee
This increase of the Planck mass replaces the expansion in the usual description. The energy density in radiation does not change with time. 
Since particle masses grow $\sim\chi$ the radiation dominated epoch will end once the energy density of matter becomes comparable to radiation. 

Comparing with model (A) with field coordinates \eqref{109F}, \eqref{109G} we find that both models coincide for $\chi^2\gg m^2$ with the same value of $\alpha$ and $\bar\lambda_c=\mu^2 M^2$. 
For $\chi^2\gg m^2$ the two models (A) and (B) are therefore equivalent, related by a simple conformal transformation. 
For a sufficient time after the end of inflation $\chi$ has grown so far that corrections $\sim m^2/\chi^2$ can be  neglected.  The cosmological solutions of models (A) and (B) become then equivalent. Nevertheless, it is instructive to understand the matter dominated epoch also in the field basis \eqref{1}, \eqref{2A} for model (B) and we will display this next.

\medskip\noindent
{\bf {\em 3. Matter domination.}}

For the subsequent matter dominated epoch the geometry remains no longer static. 
The scaling solution will correspond again to a static $\rho_m=\bar\rho$. 
However, since for matter $\rho_m$ scales $\sim\chi/a^3$, a constant $\rho$ requires the relation 
\be\label{S16}
H=\frac13\dot{s}.
\ee
For increasing $\chi$ the Universe is now expanding, although with a rate different from the usual description. 
Neglecting terms $\sim m^2/\chi^2$ and using for particle masses $\sim\chi$ the relation $T^\mu_\mu=\chi q_\chi$ the eqs. \eqref{SB3}, \eqref{SB4} become again algebraic equations,
\ba\label{SB17}
\dot{\chi}^2&=&\frac{2}{K+6}\bar\lambda_c,\nn\\
\frac{14-3K}{6}\dot{\chi}^2&=&\bar\lambda_c+\bar\rho,
\ea
with solution 
\be\label{SB18}
\frac{\bar\rho}{\bar\lambda_c}=-\frac{2(2+3K)}{3(K+6)}.
\ee
For positive $\bar\rho$ and $\bar\lambda_c+\bar\rho$ the existence of this solution requires
\be\label{SB19}
K<-\frac23~,~\alpha^2>3.
\ee

The picture of the matter dominated Universe is again rather simple. 
Both the particle masses and the scale factor increase in a way such that the energy density remains constant. 
While particle masses increase $\sim\chi\sim t$, the scale factor increases $\sim t^{1/3}$,
\be\label{139A}
H=\frac13 t^{-1}.
\ee
This differs from the expansion rate in cosmologies with a constant Planck mass. 
For both the radiation and matter dominated epoch the scaling solution includes a constant fraction of early dark energy. 
The matter dominated epoch will end if some other particle species as neutrinos have a mass that increases faster than $\chi$. 
As for model (A) this can trigger a transition to a dark energy dominated epoch. 

\medskip\noindent
{\bf {\em 4. Inflation.}}

For the inflationary epoch the intrinsic scale $m^2$ plays a role. Thus the two models (A) and (B) will yield different predictions. For model (B) two different types of scaling solutions are found in the approximation of constant $K$. Consider first the approximation $\chi^2\gg m^2$ which may become relevant towards the end of inflation. We
make the ansatz (with constant $e,f$)
\be\label{SB20}
H=f\dot{s}~,~\dot{\chi}=e,
\ee
such that eq. \eqref{SB4} reads
\be\label{SB21}
3(1+f)^2e^2=\bar\lambda_c+\frac{K+6}{2}e^2.
\ee
With $\ddot{s}=-\dot{s}^2$ eq. \eqref{SB3} yields a second quadratic equation 
\be\label{SB22}
(K+6)(1+3f)e^2=4\bar\lambda_c.
\ee
The scaling solution obeys
\ba\label{SB23}
f=\frac{K+2}{4}~,~e=4\sqrt{\frac{\bar\lambda_c}{(K+6)(3K+10)}}
\ea
and requires $K>-10/3$. It describes a linear increase of $\chi$ with time, with a decreasing Hubble parameter $H=f/t$. 

A second type of scaling solution is appropriate for the range $\chi^2\ll m^2$. Here we make the ansatz
\be\label{SB24}
H=\tilde b m~,~\dot{s}=\tilde cm
\ee
and find the algebraic expressions for the field equations
\ba\label{SB25}
K(3\tilde b\tilde c+\tilde c^2)=\frac{4\bar\lambda_c}{m^4}~,~
\tilde b^2=\frac{\bar\lambda_c}{3m^4}.
\ea
For this solution both the scale factor $a$ and the scalar field $\chi$ increase exponentially
\ba\label{SB26}
H=\sqrt{\frac{\bar\lambda_c}{3m^2}}~,~\frac{\dot{\chi}}{\chi}=\sqrt{\frac{3\bar\lambda_c}{4m^2}}
\left(\sqrt{1+\frac{16}{3K}}-1\right).
\ea
This solution exists for $K>0$ or $\tilde\alpha^2<8/3$. It is regular for $t\to -\infty$ with $\chi(t\to -\infty)\to 0$. Again, the cosmology is free of a big bang singularity.

As $\chi$ increases beyond $m$ the exponential growth \eqref{SB26} will turn over to the linear growth according to eq. \eqref{SB20}. Also the Hubble parameter does not remain constant but rather decays as $H=f/t$. 
Typical inflationary scenarios describe a transition between the two scaling solutions.
A more quantitative description of the inflationary epoch will be given after performing a Weyl scaling to the Einstein frame. 

\medskip\noindent
{\bf {\em 5. Weyl scaling.}}

For a quantitative discussion of the various cosmological epochs we perform again the Weyl scaling \eqref{W1} to the Einstein frame, resulting in the effective action \eqref{W2} with 
\ba\label{SW1}
V'&=&\frac{\bar\lambda_cM^4}{(\chi^2+m^2)^2},\nn\\
K'&=&\frac{\chi^2K}{\chi^2+m^2}+6\frac{\chi^4}{(\chi^2+m^2)^2}.
\ea
For model (B) we define $\varphi$ by
\be\label{SW2}
\varphi=\frac{M}{\alpha}\ln\frac{(\chi^2+m^2)^2}{\bar\lambda_c},
\ee
such that the exponential potential in eq. \eqref{S9} is again realized. In this convention $\chi=0$ corresponds to a minimal value $\varphi_{\rm min}=(M/\alpha)\ln (m^4/\bar\lambda_c)$. The kinetial $k^2(\varphi)$ obeys now 
\ba\label{SW3}
k^2&=&\frac{\alpha^2}{4}\left(\frac{\chi^2+m^2}{2\chi^2}\right)^2K'\nn\\
\nn\\
&=&\frac{\alpha^2}{16}\left(\frac{m^2+\chi^2}{\chi^2}K+6\right)\nn\\
\nn\\
&=&1+\alpha^2\left(\frac{1}{\tilde\alpha^2}-\frac{3}{8}\right)\frac{m^2}{\chi^2}.
\ea
It equals unity for large enough values of $\chi^2/m^2$. With the convention \eqref{SB1A} the cosmology of the radiation and matter dominated epochs is  governed by $k^2=1$. As expected, this is the same as for model (A). In particular, the 
 early dark energy fraction $\Omega_e$ is again given by eq. \eqref{14}. This holds despite the fact that in the Jordan frame the solutions have rather different characteristics.

\section{Cosmon inflation for flat cosmon potential}
\label{Cosmon inflation for flat cosmon potential}

The inflationary period is again characterized by the kinetial $k^2$ according to  eqs. \eqref{C3}, \eqref{53A}. Its form \eqref{SW3} for model (B) differs from model (A)
 for small $x=\chi^2/m^2$. 
We will find values of $\tilde\alpha$  for which $n\approx 0.95$ and $r\approx 0.05$, well compatible with present observations. On the other hand, a spectral index larger than 0.96 will lead to unacceptably high tensor fluctuations.

Defining $\bar\alpha$ by 
\be\label{SW4}
\frac{1}{\bar\alpha^2}=\frac{1}{\tilde\alpha^2}-\frac38
\ee
we infer from 
\be\label{SW5}
k^2=1+\frac{\alpha^2}{\bar\alpha^2x}
\ee
the slow roll parameters
\ba\label{SW6}
\epsilon&=&\frac{\bar\alpha^2x}{2}
\left(1+\frac{\bar\alpha^2x}{\alpha^2}\right)^{-1},\\
\eta&=&2\epsilon-\frac{1+x}{2}\frac{\partial\epsilon}{\partial x}\nn\\
&=&\frac{\bar\alpha^2}{4\left(1+\frac{\bar\alpha^2x}{\alpha^2}\right)}
\left(-1+3x+\frac{\bar\alpha^2x(1+x)}{\alpha^2\left(1+\frac{\bar\alpha^2 x}{\alpha^2}\right)}
\right).\nn
\ea
The relation between $x$ and the number of $e$-folding before the end of inflation at $x_f$ is given by
\ba\label{SW7}
&&N(x)=\frac{2}{\alpha^2}\int^{x_f}_{x}dx
\frac{k^2(x)}{1+x}\nn\\
&&=\frac{2}{\bar\alpha^2}
\int^{x_f}_x dx\left(\frac{1}{x(1+x)}+\frac{\bar\alpha^2}{\alpha^2(1+x)}\right).
\ea
We observe that $\eta$ becomes negative for $x\to0$, corresponding to the concave region of $V'$ in eq. \eqref{SW1}. 

With 
\ba\label{SW8}
N(x)=\frac{2}{\bar\alpha^2}\ln\left(\frac{x_f}{x}\right)-
\left(\frac{2}{\A}-\frac{2}{\alpha^2}\right)\ln
\left(\frac{1+x_f}{1+x}\right),
\ea
and $\epsilon(x_f)=1$, 
\be\label{SW9}
x_f=\frac{1}{2\bar\alpha^2\left(1-\frac{1}{2\alpha^2}\right)}\approx
\frac{1}{2\bar\alpha^2},
\ee
we can now compute $\epsilon(N)$ and $\eta(N)$. The corresponding values of $n$ and $r$ are given in Table II for different values of $\tilde\alpha$. There the first number refers to $N=60$, and the second number in brackets to $N=50$. 
\begin{table}[h!]\label{Table2}
\centering
\begin{tabular}{|p{10mm}|p{20mm}|p{20mm}|p{20mm}|}\hline
\centering{$\tilde\alpha$}&\centering{$0.24$}&\centering{$0.28$}&\centering{$0.325$}\tabularnewline\hline
\centering{$n$}&\centering{$0.954~(0.95)$}&\centering{$0.95~(0.944)$}&\centering{$0.94~(0.936)$}\tabularnewline\hline
\centering{$r$}&\centering{$0.08~(0.12)$}&\centering{$0.054~(0.085)$}&\centering{$0.027~(0.049)$}\tabularnewline\hline
\centering{$\frac {m}{(\bar\lambda_c)^{1/4}}$}&
\centering{$129~(114)$}&\centering{$150~(131)$}&\centering{$182~(156)$}\tabularnewline\hline
\end{tabular}
\caption{Properties of density fluctuations, model (B).}
\end{table}

We may again distinguish between the regime of large $x$ at the time of horizon crossing of density perturbations and the small-$x$ regime. For $x\gg1$ one has
\ba\label{SW10}
N(x)&\approx&
\left(\frac{2}{\A}-\frac{2}{\alpha^2}\right)
\left(\frac1x-2\A\right)
-\frac{2}{\alpha^2}\ln(2\A x)\nn\\\nn\\
&\approx&\frac{2}{\A x}-4,
\ea
and therefore
\ba\label{SW11}
\epsilon&=&\frac{\A x}{2}=\frac{1}{N+4},\nn\\
\eta&=&\frac32\epsilon=\frac{3}{2(N+4)}.
\ea
The resulting large tensor component $r\approx 0.25$ is disfavored by observation. 

On the other side, for $x\ll 1$ one finds
\ba\label{SW12}
N&=&\frac{2}{\A}\left[\ln\left(\frac 1x\right)-\ln(1+2\A)+x\right]\nn\\\nn\\
&\approx&\frac{2}{\A}\ln\left(\frac 1x\right).
\ea
This yields
\ba\label{SW13}
\epsilon&\approx&\frac{\A}{2}\exp\left(-\frac{\A N}{2}\right),\nn\\
\eta&\approx&-\frac{\A}{4}\left(1-3\exp\left(-\frac{\A N}{2}\right)\right),\nn\\
x&\approx&\exp\left(-\frac{\A N}{2}\right),
\ea
and therefore
\ba\label{SW14}
n&=&1-\frac{\A}{2}(1+3x),\nn\\
r&=&8\A x.
\ea
For $\bar\alpha^2$ substantially larger than $2/N$ the horizon crossing of characteristic fluctuations occurs at very small $x$ and therefore leads to very small $r$, 
\ba\label{SW15}
r&=&8\A\exp\left(-\frac{\A N}{2}\right),\nn\\\nn\\
n&=&1-\frac{\A}{2}-\frac{3r}{16}.
\ea

Since both $r$ and $n$ are determined by $\A$ we can establish a relation between these two quantities. An approximate form for small $r$ reads 
\ba\label{SW16}
r=\frac{16(1-n)\exp\big(-N(1-n)\big)}{1-3\big[N(1-n)-1\big]\exp\big(-N(1-n)\big)}.
\ea
For $n=0.95$ this yields the prediction $(N=60)$ 
\be\label{SW17}
r=0.05.
\ee
This model of cosmon inflation seems to be compatible with all present data. The predicted amplitude of tensor fluctuations may be detectable in the data of the Planck satellite. We can interpret the measurement of the spectral index as a measurement of $\tilde\alpha$. For $n=0.95$ one finds
\be\label{SW18}
\A=0.08,\ \tilde\alpha=0.28.
\ee
For model (B) a value of $n$ larger than $0.96$ would require an unacceptable high power of tensor fluctuations.

We finally determine the mass scale $m$ from the observed amplitude of the density fluctuations
\be\label{SW20}
24\pi^2\Delta^2=\frac{V'}{\epsilon M^4}\approx 5\cdot 10^{-7},
\ee
with 
\be\label{SW21}
\frac{V'}{M^4}=\frac{\bar\lambda_c}{m^4}(1+x)^{-2}.
\ee
The corresponding values of $m$ are displayed in table II. They are about two orders of magnitude larger than the characteristic scale for the potential $(\bar\lambda_c)^{1/4}$. With the numerical value \eqref{SB2} one obtains values $m\approx 0.3$ eV.

\section{Conclusions and discussion}
\label{Conclusions}

We have found that models of gravity coupled to a single scalar field can connect inflation with the present Dark Energy dominated epoch. It is remarkable that the same simple potential for the cosmon can describe the whole history of Dark Energy - from its domination during inflation to a subleading Early Dark Energy during radiation and matter domination, and finally again to domination in the present Universe. This evolution spans 60 orders of magnitude in time or 120 orders of magnitude in the ratio between the potential and the fourth power of the effective Planck mass.

It may be expected that by using arbitrary functions $F,K$ and $V$ in eq. \eqref{1} such a unified description of inflation and present Dark Energy becomes possible. This extends to other (equivalent) formulations of a scalar degree of freedom coupled to the graviton as $f(R)$-theories. The novelty of our approach concerns the simplicity of our models. The scalar potential $V$ involves only one mass parameter, $\mu=2\cdot 10^{-33}$eV for model (A) and $(\bar\lambda_c)^{1/4}=2\cdot 10^{-3}$eV for model (B). No free dimensionless couplings appear. The effective Planck mass being a dynamical variable there is no fixed mass parameter for it. Thus no dimensionless parameters have to be tuned to render the tiny observed ratio between the present Dark Energy density and the fourth power of the present Planck mass. 

The mass parameter in the potential sets the typical time scale for the evolution of the universe. For model (B) the Hubble parameter never exceeds the eV-region. Even more striking, for model (A) the Hubble parameter remains of the order of the present Hubble parameter $H_0=0.69\mu$ for all cosmological epochs, including inflation and the approach to the ``big bang''! In contrast to the diverging Hubble parameter for the standard big bang picture we deal here with a ``slow universe''. For all times from minus infinity to plus infinity the characteristic time scale of the cosmic evolution is of the order $10^{10}$ yr. This time scale also governs the evolution of particle masses. 

Our models are characterized by two different asymptotic regions: very early cosmology corresponds to the limit $\chi\to 0$, and late cosmology is characterized by $\chi\to\infty$. The transition between the asymptotic regions involves a mass scale $m$, such that we can form a dimensionless quantity $x=\chi^2/m^2$. Early cosmology corresponds to $x\ll 1$, late cosmology to $x\gg 1$. The transition scale $m$ introduces a dimensionless parameter $m/\mu$ (A) or $m(\bar\lambda_c)^{-1/4}$ (B). Its value can be fixed by the amplitude of the primordial density fluctuations. For model (B) the function $F$ for the scalar-gravity coupling involves no parameter besides $m$. There is no parameter at all for model (A).

The kinetial $K$ multiplies the derivative term for the cosmon in the effective action. It plays a similar role as the wave function renormalization, with the particularity that stability requires $K>-6$  if the effective Planck mass is proportional to the cosmon field $\chi$. (For a fixed Planck mass the stability bound would be $K>0$.) Compatibility with the observed properties of the density fluctuations generated during inflation requires $K$ to take a large positive value during inflation. On the other  hand it has to be close to the conformal value $-6$ in the later stages of the cosmic evolution in order  to keep the amount of Early Dark Energy small. We therefore introduce two dimensionless parameters $K_0=K(\chi\to 0)$ and $K_\infty=K(\chi\to\infty)$. In model (A) they are connected to the parameters $\alpha$ and $\tilde \alpha$ by $K_0+6=4/\tilde\alpha^2,~K_\infty+6=4/\alpha^2$, while there is an additional factor four for model (B). The sector of the scalar field coupled to gravity is therefore described by only three dimensionless parameters $\alpha,\tilde \alpha$ and $m/\mu$ or $m(\bar\lambda_c)^{-1/4}$. For definiteness we have chosen a specific form \eqref{2}, \eqref{SB1A} for $K(\chi)$. The precise shape of the transition between $K_0$ and $K_\infty$ at $x\approx 1$ is not very important for a description of realistic cosmology.

Additional couplings appear in the matter sector. Our models assume that for large $\chi$ the masses of all particles except neutrinos are given by $m_i=h_i\chi$. The values of the dimensionless couplings $h_i$ are determined by the present particle masses $m^{(0)}_i$ in units of the Planck mass $M,h_i=m^{(0)}_i/M$. They therefore do not involve new parameters besides the standard fixed  particle  masses in the conventional big bang picture. We do not need to specify the couplings $h_i$ for small $\chi$ since particles play no role in the very early cosmological epochs. It is well conceivable that these couplings differ for early epochs from the present values. This may play a role for the entropy production after inflation. We have left this interesting topic out of the scope of the present work. 

An important ingredient of our models is the assumption that neutrino masses do not scale $\sim \chi$, due to a mass parameter in the sector of heavy singlets (of the standard model gauge group) that decreases rather than to scale $\sim\chi$. This mass parameter enters the light neutrino masses by some type of seesaw mechanism. The resulting increase of the neutrino masses faster than $\chi$ stops the evolution of the cosmon once neutrinos become non-relativistic, thereby triggering the onset of the accelerated expansion. The crucial parameter in the neutrino sector is the (present) effective growth parameter $\tilde \gamma$, as determined by $m_\nu\sim\chi^{2\tilde\gamma+1}$. This parameter enters the striking relation \eqref{52J} which determines the present Dark Energy density in terms of the neutrino mass. The observation of Dark Energy yields
\be\label{CX1}
\tilde\gamma=6.15\left(\frac{m^{(0)}_\nu}{{\rm eV}}\right)^{-1},
\ee
with $m^{(0)}_\nu$ the present average neutrino mass. The parameter $\tilde\gamma$ may be considered as the equivalent of $\Lambda/M^4\approx 10^{-120}$ in the $\Lambda$CDM-model. While it needs to be fixed (for given $m^{(0)}_\nu$) in order to account for the present fraction of $70$\% Dark Energy, it is a quantity roughly of order one that does not need fine tuning. (Depending on the details of the neutrino sector $\tilde\gamma$ may actually be a growing function of $\ln(\chi/m)$.)

Due to its few parameters our models are subject to many observational tests. Let us compare our model (A) with the the standard $\Lambda$CDM-model. The parameter $\tilde\gamma$ corresponds to the parameter $\Lambda$, fixing the present Dark Energy density or $\Omega_m$ (assuming $\Omega_{{\rm tot}}=1)$. The parameter $m/\mu$ is fixed by the amplitude of the primordial density fluctuations which is also a free parameter in the $\Lambda$CDM-model. The spectral index $n$ of the primordial fluctuations depends weakly on $\tilde \alpha$. This may be used to fix this parameter. Then our model has only one additional parameter, namely $\alpha$, which determines the fraction of Early Dark Energy. This is subject to observational tests which already constrain this parameter to $\alpha\gtrsim 10$. 

With all parameters fixed in this way our model leads to several testable predictions. The tensor to scalar ratio $r$ is computable. The predictions of $(n,r)$ in table I make the model falsifiable in the sense that an 
additional parameter, as a constant in $V(\chi)$, may be needed. (For the time being the estimate of the allowed parameter range $(n,r)$ has to wait until the non-standard neutrino sector is included in the analyses.) Also non-gaussianities or other features beyond the simple single field inflation may make extensions of the model necessary. Concerning the present Dark Energy, our models predict an equation of state $w$ rather close to $-1$, but somewhat above. More striking, the predicted formation of large scale neutrino lumps renders the cosmic neutrino background observable. Furthermore, the large coupling between neutrinos and Dark Energy will lead to deviations of the present value of the Hubble parameter from the CMB-inferred value for the $\Lambda$CDM-model.

The evolution of the Universe  after the inflationary epoch can be understood as the approach to a fixed point. 
This is realized by a cosmological ``runaway solution'' where $\chi$ increases continuously. Dimensionless couplings or mass ratios are, in general, functions of $\chi$. An asymptotic fixed point means that these quantities become independent of $\chi$ for $\chi\to\infty$.
For a fixed point dilatation symmetry becomes exact - the memory of all intrinsic mass scales is lost. All particle masses and the Planck mass scale   proportional  to $\chi$. The observed non-zero masses are therefore connected to a spontaneous  breaking of dilatation symmetry. The corresponding  Goldstone boson corresponds  to the cosmon which becomes massless for $\chi\to \infty$. At the present time the Universe has not yet reached the fixed point. Residual scaling violation in the effective action is responsible for the present tiny Dark Energy density. 

Dilatation symmetry is most easily visible in the Jordan frame where scale transformations act multiplicatively on all fields. In the Jordan frame all explicit mass parameters reflect a violation of dilatation symmetry. For this reason we have presented a comprehensive discussion of all cosmological epochs in the Jordan frame, even though easy contact to the standard cosmological  observables is facilitated in the Einstein frame. Furthermore, the cosmology in the Jordan frame is free of a big bang singularity.

For observations the new aspects of the present investigation, beyond tests of existing models of growing neutrino quintessence \cite{ABW,CWNEU,GNQ1,GNQ2,GNQ3,GNQ4,GNQ5,GNQ5A,GNQ7,GNQ8,GNQ9,GNQ10} or Early Dark Energy, concern the predictions for the amplitude and shape of the density fluctuations generated during inflation. The unified description of ``primordial and Dark Energy'' raises the interesting question if observable properties of the inflationary epoch can be linked to observations of the present Dark Energy. Unfortunately, the parameters $\alpha$ and $\tilde\gamma$ determine the properties of late Dark Energy (quintessence), but play no role for primordial Dark Energy (inflation). In turn, $\tilde \alpha$ and $m$ determine the primordial fluctuations, but do not show up in late cosmology. Only the mass scale $\mu$ (A) or $(\bar\lambda_c)^{1/4}$ (B) plays a crucial role both of primordial and late Dark Energy. It is not a dimensionless parameter, however, and cannot be used for relating predictions for early and late cosmology. 

The two examples of simple models presented here demonstrate that a rather wide class of inflationary models can be realized within the framework of ``cosmon inflation''. Generalizations are straightforward. One example uses 
\be\label{211A}
F=m^2+\chi^2~,~V=\bar\lambda_c+\mu^2\chi^2,
\ee
another may employ a modified form of the shape of the kinetial. For late cosmology only the leading term for large $\chi$ matters, disconnecting again the predictions for early and late cosmology. Late cosmology for the model \eqref{211A} corresponds to model (A), such that we may view this setting as an extension of model (A) which matters for the inflationary period. It modifies the predictions $(n,r)$ for the primordial fluctuation spectrum, for example $n=0.96, r=0.04$ for $\tilde \alpha=0.18$ \cite{CI}.

One may ask if a closer connection between primordial and late Dark Energy may be found if other degrees of freedom as the Higgs scalar are taken into account. In this paper we have concentrated on a single scalar  field. This is sufficient for a description of the overall cosmology. On the other hand, it is clear that additional scalar fields are needed for a realistic scenario of particle physics. In particular, the Higgs doublet  $\tilde h$ is responsible for the masses of the charged particles  in the standard model. Furthermore, it seems  very likely that other fields besides the cosmon play an important role for the entropy production and heating at the end of inflation. This is the reason why we have left  out this subject in the present paper. (For a short discussion see ref. \cite{CI}.)

At this place we only comment that a minimal setting where the Higgs doublet mediates the entropy production can be realized within the general framework discussed here. We may write the interactions between the Higgs doublet and the cosmon as

\be\label{187A}
\tilde V_h=\frac{1}{2}\lambda_h(\tilde h^\dagger\tilde h)^2+\lambda_\chi\chi^4+\gamma\tilde h^\dagger\tilde h\chi^2,
\ee
with dimensionless functions $\lambda_h,\ \lambda_\chi$ and $\gamma$ depending on $x=\chi^2/m^2$. For $x\to\infty$ these functions  should approach the constants specified by eq. \eqref{C1a}, corresponding to a very small negative $\gamma=-\varepsilon_h \lambda_h$ and $\lambda_\chi$ fixed in terms of $\lambda_h$ and $\gamma$.
Not much is known about these functions for the values of $x$ characteristic for the inflationary period. For example, one could imagine  that $\gamma$ is positive for small $x$, turns to a substantial negative value at the end of inflation and finally settles at a tiny negative value for $x\to\infty$.  In the early stages of inflation the Higgs doublet could perform small oscillations around $\tilde h=0$, leaving potentially a periodic imprint in the spectrum of primordial density fluctuations \cite{Peiris}. Once $\gamma$ turns negative the value $\tilde h=0$ becomes unstable. Substantial oscillations around the new minimum for $\tilde h$ could produce incoherent particles of all species coupling to $\tilde h$. This would result in efficient entropy production. Other fields besides the Higgs doublet could take this role as well.

Again, possible direct links between predictions of observable quantities during inflation and properties of the Higgs scalar are obscured by our lack of knowledge of the cosmon potential and kinetial and its couplings to the Higgs field or other relevant fields.

One new situation may arise if it becomes possible to compute the fluctuations $F,K$ and $V$. A first attempt in this direction are functional renormalization group computations in dilaton quantum gravity \cite{HPRW}. If it is possible to establish the suggested fixed point (scaling solution) this would not only constitute a viable candidate for a non-perturbatively renormalizable quantum field theory for gravity. It would also provide the fixed point towards which cosmology converges for $\chi\to\infty$. For a given fixed point also the deviations from the fixed point are often computable, which concerns in our context the role  of the explicit mass scales $m,\ \mu$ or $ (\bar\lambda_c)^{1/4}$. The functions $F,K$ and $V$ would become computable. In this event the unified picture sketched in the present paper could develop its full power, relating observables for the primordial and late time cosmology.

\bigskip\noindent

\bigskip\noindent
{\em Acknowledgment}

\noindent
Part of this work was performed at KITP, Santa Barbara, and the author thanks for hospitality.
This research was supported by the National Science Foundation under Grant No. NSF PHY11-25915.

\bibliography{variable_gravity_universe}

\end{document}